\documentclass[aps,prd,reprint,groupedaddress,showpacs, superscriptaddress]{revtex4-1}

\usepackage{amsmath}
\usepackage{graphicx,color}
\usepackage[colorlinks=true, linkcolor=blue, citecolor=blue, linktoc=all]{hyperref}

\begin{document}

\title{The Sensitivity of the Advanced LIGO Detectors at the Beginning of Gravitational Wave Astronomy}

\author{%
D.~V.~Martynov,$^{1}$  
E.~D.~Hall,$^{1}$  
B.~P.~Abbott,$^{1}$  
R.~Abbott,$^{1}$  
T.~D.~Abbott,$^{2}$  
C.~Adams,$^{3}$  
R.~X.~Adhikari,$^{1}$  
R.~A.~Anderson,$^{1}$    
S.~B.~Anderson,$^{1}$  
K.~Arai,$^{1}$
% ...
\\
\medskip
}\noaffiliation
\affiliation {LIGO, California Institute of Technology, Pasadena, CA 91125, USA }
\affiliation {Louisiana State University, Baton Rouge, LA 70803, USA }
\affiliation {LIGO Livingston Observatory, Livingston, LA 70754, USA }
\affiliation {University of Florida, Gainesville, FL 32611, USA }
\affiliation {Syracuse University, Syracuse, NY 13244, USA }
\affiliation {LIGO Hanford Observatory, Richland, WA 99352, USA }
\affiliation {SUPA, University of Glasgow, Glasgow G12 8QQ, United Kingdom }
\affiliation {LIGO, Massachusetts Institute of Technology, Cambridge, MA 02139, USA }
\affiliation {Columbia University, New York, NY 10027, USA }
\affiliation {University of Western Australia, Crawley, Western Australia 6009, Australia }
\affiliation {Albert-Einstein-Institut, Max-Planck-Institut f\"ur Gravi\-ta\-tions\-physik, D-30167 Hannover, Germany }
\affiliation {Stanford University, Stanford, CA 94305, USA }
\affiliation {Leibniz Universit\"at Hannover, D-30167 Hannover, Germany }
\affiliation {The University of Sheffield, Sheffield S10 2TN, United Kingdom }
\affiliation {University of Sannio at Benevento, I-82100 Benevento, Italy and INFN, Sezione di Napoli, I-80100 Napoli, Italy }
\affiliation {The University of Mississippi, University, MS 38677, USA }
\affiliation {Rutherford Appleton Laboratory, HSIC, Chilton, Didcot, Oxon OX11 0QX, United Kingdom }
\affiliation {University of Michigan, Ann Arbor, MI 48109, USA }
\affiliation {American University, Washington, D.C. 20016, USA }
\affiliation {University of Massachusetts-Amherst, Amherst, MA 01003, USA }
\affiliation {University of Oregon, Eugene, OR 97403, USA }
\affiliation {University of Adelaide, Adelaide, South Australia 5005, Australia }
\affiliation {SUPA, University of Strathclyde, Glasgow G1 1XQ, United Kingdom }
\affiliation {Australian National University, Canberra, Australian Capital Territory 0200, Australia }
\affiliation {University of Minnesota, Minneapolis, MN 55455, USA }
\affiliation {Hobart and William Smith Colleges, Geneva, NY 14456, USA }
\affiliation {Faculty of Physics, Lomonosov Moscow State University, Moscow 119991, Russia }
\affiliation {The University of Texas Rio Grande Valley, Brownsville, TX 78520, USA }
\affiliation {MTA E\"otv\"os University, ``Lendulet'' Astrophysics Research Group, Budapest 1117, Hungary }
\affiliation {SUPA, University of the West of Scotland, Paisley PA1 2BE, United Kingdom }
\affiliation {University of Maryland, College Park, MD 20742, USA }
\affiliation {California State University Fullerton, Fullerton, CA 92831, USA }
\affiliation {Monash University, Victoria 3800, Australia }
\affiliation {University of Birmingham, Birmingham B15 2TT, United Kingdom }
\affiliation {University of Washington, Seattle, WA 98195, USA }
\date{\today}

\begin{abstract}
The Laser Interferometer Gravitational Wave Observatory (LIGO) consists of two widely separated 4\,km laser interferometers designed to detect gravitational waves from distant astrophysical sources in the frequency range from 10\,Hz to 10\,kHz.
The first observation run of the Advanced LIGO detectors started in September 2015 and ended in January 2016.
A strain sensitivity of better than $10^{-23}/\sqrt{\text{Hz}}$ was achieved around 100\,Hz.
Understanding both the fundamental and the technical noise sources was critical
for increasing the astrophsyical strain sensitivity.
The average distance at which coalescing binary black hole systems with individual masses of 30\,$M_\odot$ could be detected above a signal-to-noise ratio (SNR) of 8 was 1.3\,Gpc, and the range for binary neutron
star inspirals was about 75\,Mpc.
With respect to the initial detectors, the observable volume of the Universe
increased by a factor 69 and 43, respectively.
These improvements helped Advanced LIGO to detect the gravitational wave signal from the binary black hole coalescence, known as GW150914.
\end{abstract}


\maketitle

\section{Introduction}
\label{sec:intro}

The possibility of using interferometers as gravitational wave detectors was first considered in the early 1960s \cite{intro_ger}. In the 1970s and 1980s, long-baseline broadband laser interferometric detectors were proposed with the potential for an astrophysically interesting sensitivity ~\cite{weiss:1972, intro_drever}. Over several decades, this vision evolved into a world-wide network of ground based interferometers~\cite{intro_ligo, intro_virgo, intro_GEO}. These instruments target gravitational waves produced by compact binary coalescences, supernovae, non-axisymmetric pulsars, cosmological background as well as any unknown astrophysical sources in the audio frequency band, from 10\,Hz to 10\,kHz~\cite{riles_2013}.

The first generation of LIGO detectors consisted of two 4-km-long and one 2-km-long interferometers in the United States~\cite{intro_ligo_2009}: L1 in Livingston, Louisiana, H1 and H2 in Hanford, Washington. They were operational until 2010 and reached their designed strain sensitivity over the detection band, with a peak sensitivity of $2 \times 10^{-23} /\sqrt{\mathrm{Hz}}$ at 200\,Hz. Astrophysically relevant results were produced by the initial LIGO detectors \cite{intro_iligo_stochastic, intro_iligo_cw, intro_iligo_cbc, intro_iligo_bursts}, however, no gravitational wave signals were detected.

The second generation Advanced LIGO detectors~\cite{design_aligo_peter} were installed in the existing facilities from 2010 to 2014. This new generation of instruments was designed to be 10~times more sensitive than initial LIGO, and promised to increase the volume of the observable universe by a factor of 1000. Commissioning of the newly$-$installed detectors took place from mid 2014 to mid 2015. In September 2015, Advanced LIGO began the era of gravitational wave astronomy with its first observation run (O1), collecting data until January 2016. 
This run has culminated in the first direct detection of gravitational waves from the black hole coalescence, GW150914~\cite{results:2016,astro:2016}. This system consisted of two black holes of about 35 solar mass each which merged about 500\,Mpc away. 

While the detectors were not yet operating at design sensitivity during the first observation run, their astrophysical reach was already significantly greater than that of any previous detector in the frequency range 10\,Hz--10\,kHz. 
Around 100\,Hz, the strain sensitivity was $8 \times 10^{-24}/\sqrt{\mathrm{Hz}}$. 
For a system consisting of two 30\,$M_\odot$ black holes the sky location and source orientation-averaged range was 1.3\,Gpc, whereas for a binary neutron star system the range was 70--80\,Mpc.
This range is $\simeq$4.1 and $\simeq$3.5 times higher than that of the initial LIGO detectors, resulting in a factor of $\simeq$70 and $\simeq$40 improvement, respectively, of the volume that is probed and LIGO's detection potential.

In this paper we describe the noise characterization of the Advanced LIGO detectors during the first observation run. Sec.~\ref{sec:ifoconf} introduces the optical configuration, control system and calibration of the detectors. Sec.~\ref{sec:instnoise} analyzes the performance of the detectors and describes all investigated noise sources. We end with the conclusions in Sec.~\ref{sec:conclusions}.

\begin{figure}[ht!]
	\centering
	\includegraphics[width=0.45\textwidth]{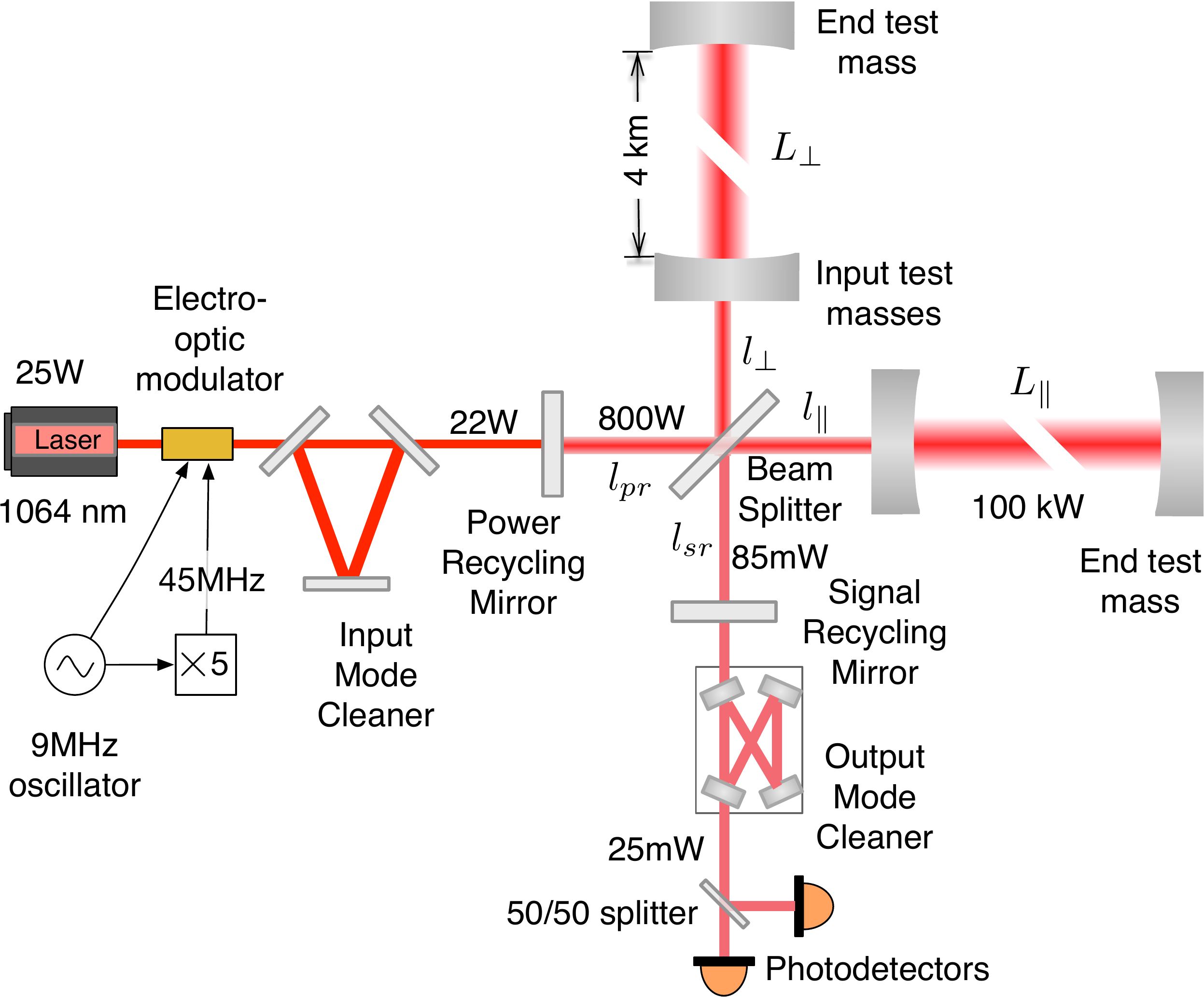}
    \caption[Layout of Advanced LIGO]{Layout of an Advanced LIGO detector. The annotations show the optical power in use during O1. These power levels are a factor of $\simeq$8 smaller compared to the designed power levels. The Nd:YAG laser~\cite{design_psl}, with wavelength $\lambda$=1064\,nm, is capable of producing up to 180\,W, but only 22\,W were used. A suspended, triangular Fabry-Perot cavity serves as an input mode cleaner~\cite{chris_thesis, mueller_ioo_2016} to clean up the spatial profile of the laser beam, suppress input beam jitter, clean polarization, and to help stabilize the laser frequency. The Michelson interferometer is enhanced by two 4-km-long resonant arm cavities, which increase the optical power in the arms by a factor of $G_{\text{arm}} \simeq 270$. Since the Michelson interferometer is operated near a dark fringe, all but a small fraction of the light is directed back towards the laser. The power recycling mirror resonates this light again to increase the power incident on the beamsplitter by a factor of $\simeq 40$, improving the shot noise sensing limit and filtering laser noises. On the antisymmetric side, the signal recycling mirror is used to broaden the response of the detector beyond the linewidth of the arm cavities. An output mode cleaner is present at the antisymmetric port, to reject unwanted spatial and frequency components of the light, before the signal is detected by the main photodetectors.}
	\label{fig:detector}
\end{figure}

\begin{figure}[tbh]
    \begin{center}
    \includegraphics[width=0.45\textwidth]{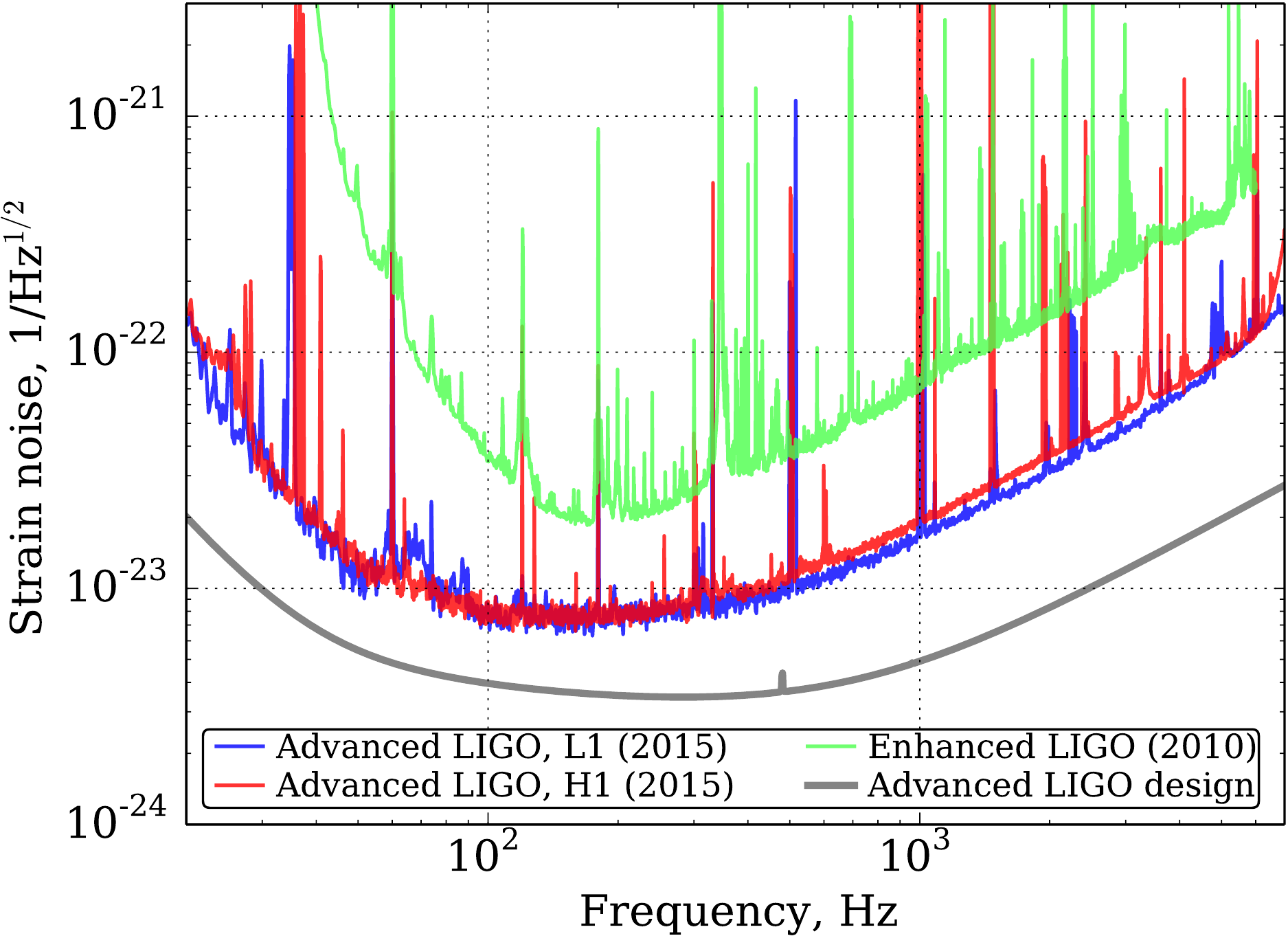}
    \caption[Strain Sensitivity]{The strain sensitivity for the LIGO Livingston detector (L1) and the LIGO Hanford detector (H1) during O1. Also shown is the noise level for the Advanced LIGO design (gray curve) and the sensitivity during the final data collection run (S6) of the initial detectors.}
    \label{fig:sensitivity}
    \end{center}
\end{figure}

\begin{figure}[tbh]
    \begin{center}
    \includegraphics[width=0.45\textwidth]{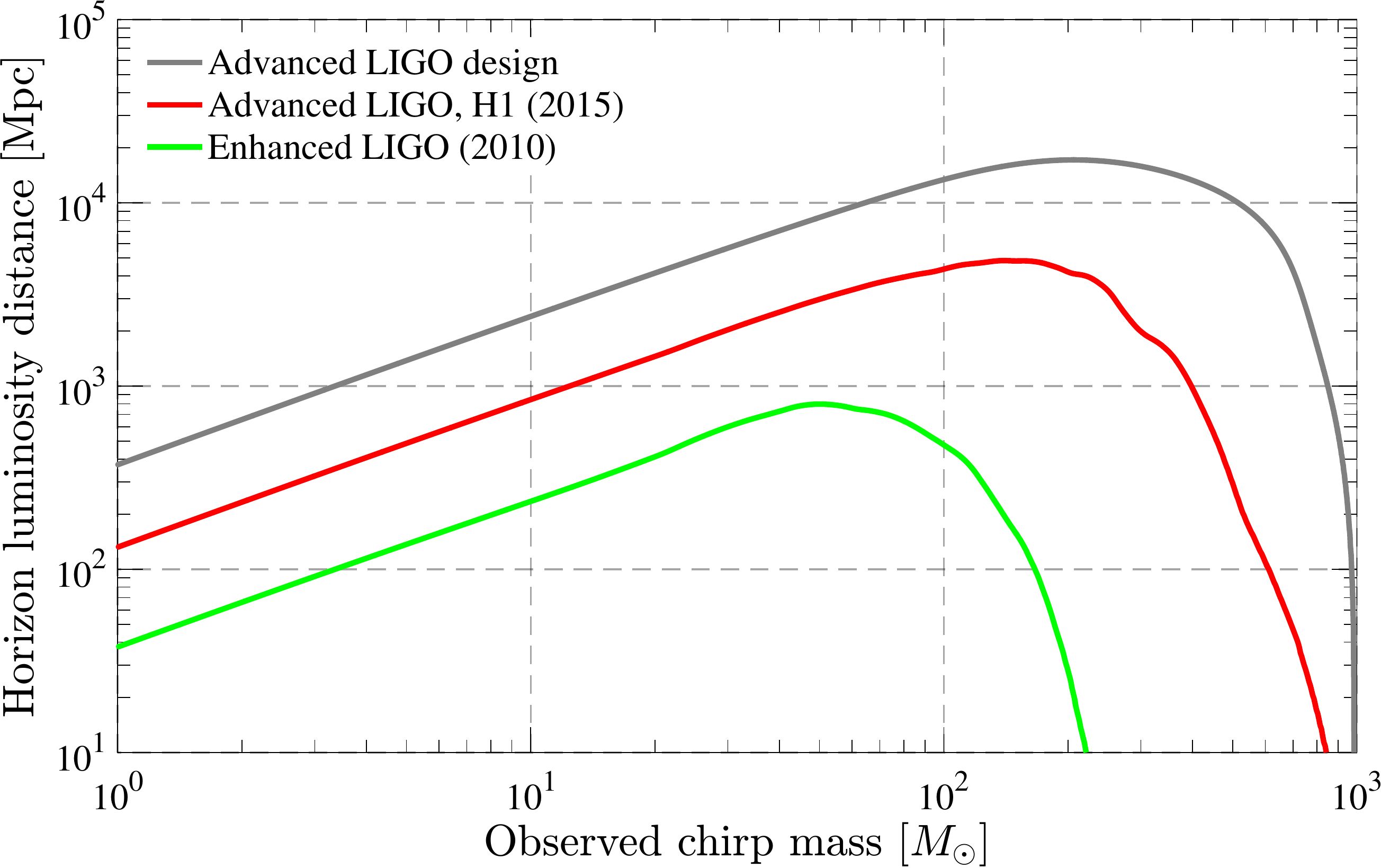}
    \caption[Binary Inspiral Range]{The sensitivity to coalescing compact binaries for the Advanced LIGO design, first observation run (O1) and the final run with the initial detectors (S6). The traces show the horizon distance, which is the distance along the most sensitive direction of the interferometer for a binary inspiral system that is seen head-on and for a signal-to-noise ratio of 8. The horizontal axis is the chirp mass which is defined as $\mathcal{M} = (1+z) \mu^\frac35 M^\frac25$, where $M = M_1 + M_2$ is the total mass, $\mu = M_1 M_2 / M$ is the reduced mass, and $z$ is the cosmological redshift. Units are in solar masses, $M_\odot$. The horizon distance is computed for the case of equal masses $M_1$ = $M_2$ and using the inspiral--merger model from~\cite{merger_b}.}
    \label{fig:range}
    \end{center}
\end{figure}

\section{Interferometer Configuration}
\label{sec:ifoconf}

In general relativity, a gravitational wave far away from the source can be approximated as a linear disturbance of the Minkowski metric, $g_{\mu\nu} = \eta_{\mu\nu} + h_{\mu\nu}$ with the space-time deformation expressed as a dimensionless strain, $h_{\mu\nu}$. In a Michelson interferometer we define the differential displacement as $L = L_{\parallel} - L_{\bot}$, where $L_{\parallel}$ and $L_{\bot}$ are the lengths of the inline arm and the perpendicular arm, respectively, as shown in Fig.~\ref{fig:detector}. With equal macroscopic arm lengths, $L_0 \simeq L_{\parallel}\simeq L_{\bot}$, the gravitational wave strain and the differential arm length are related through the simple equation $L(f) = L_{\parallel}-L_{\bot}=h(f) L_0$, where $h$ is the average differential strain induced into both arms at frequency $f$.

The test masses are four suspended mirrors that form Fabry-Perot arm cavities. These mirrors can be considered as inertial masses above the pendulum resonance frequency ($\sim$1\,Hz). Any noise present in the differential arm channel is indistinguishable from a gravitational wave signal. Residual seismic noise, thermal noise associated with the vertical suspension resonance, and the gravity-gradient background limits the useful frequency range to above 10\,Hz as discussed in Sec.~\ref{seismic_thermal}. Motion of the four test masses form the two most relevant degrees of freedom: differential and common arm lengths. While gravitational waves couple to the differential arm length, the common arm length is highly sensitive to changes in the laser frequency according to the equation
\begin{equation}
L_+(f) = \frac{L_{\parallel}+L_{\bot}}{2} = L_0 \frac{V(f)}{\nu},
\end{equation}
where $\nu=2.82\times10^{14}$\,Hz is the laser carrier frequency, $V(f)$ is the laser frequency noise. Signal $L_+$ is used for frequency stabilization of the main laser as discussed in Sec.~\ref{noises_laser}.

The central part of the interferometer is usually called the dual-recycled Michelson interferometer. Its function is to optimize the detector's response to gravitational waves. The power recycling cavity, formed by the power recycling mirror and the two input test masses, increases optical power incident on the arm cavities and passively filters laser noises as discussed in Sec.~\ref{noises_laser}.  The signal recycling cavity, formed by the signal recycling mirror and the two input test masses, is used to broaden the response of the detector beyond the linewidth of the arm cavities. The Michelson interferometer, formed by the beam splitter and the two input test masses, is controlled to keep the antisymmetric port near the dark fringe. The dual recycled Michelson interferometer can thus be described by three degrees of freedom: power recycling cavity length $l_{p,+}$, signal recycling cavity length $l_{s,+}$ and Michelson length $l_-$, defined as
\begin{equation}
\begin{split}
	l_{p,+} & = l_{\text{pr}} + \frac{l_{\parallel} + l_{\bot}}{2} \\
	l_{s,+} & = l_{\text{sr}} + \frac{l_{\parallel} + l_{\bot}}{2} \\
	l_{-} & = l_{\parallel} - l_{\bot},
\end{split}
\end{equation}
where distances $l_{pr}, l_{sr}, l_{\parallel}$ and $l_{\bot}$ are defined in Fig.~\ref{fig:detector}. 

The most important optical parameters of the Advanced LIGO interferometers are summarized in Table~\ref{table:optparam}. The beam size here is defined as the distance from the beam center to the point when intensity is reduced by a factor $1/e^2$. The cavity pole $f_p$ determines the width of the cavity resonance and is given by
\begin{equation}
	f_p = \frac{Y c} {8 \pi L_0},
\end{equation}
where $c$ is the speed of light and $Y \ll 1$ is the total optical loss in the cavity, including transmission of the input and output cavity couplers as well as scattering and absorption  losses. The response of the Advanced LIGO interferometers is diminished at high frequencies due to common and differential coupled cavity poles ($f_+$ and $f_-$) according to the transfer functions
\begin{equation}\label{eq:poles}
	K_{+} = \frac{f_+}{if + f_+}; \hspace{0.5cm}
	K_{-} = \frac{f_-}{if + f_-}.
\end{equation}

\begin{table}[h!]
    \caption{List of optical parameters}\label{table:optparam}
    \begin{ruledtabular}
    \begin{tabular}{l r r}
    Parameter & Value & Unit \\ \hline
    Laser wavelength & 1064 & nm \\
    Arm cavity length, $L_0$ & 3994.5 & m \\
    Power recycling cavity length, $l_{p,+}$ & 57.66 & m \\
    Signal recycling cavity length, $l_{s,+}$ & 56.01 & m \\
    Michelson asymmetry, $l_-$ & 8 & cm \\
    Input mode cleaner length (round trip) & 32.95 & m \\
    Output mode cleaner length (round trip) & 1.13 & m \\
    Input mode cleaner finesse & 500 & \\
    Output mode cleaner finesse & 390 & \\
    Round trip loss in arm cavity, $Y_{\text{arm}}$ & 85--100 & ppm \\
    Arm cavity build--up, $G_{\text{arm}}$ & 270 & \\
    Power recycling gain, $G_{\text{prc}}$ & 38 & \\
    Signal recycling attenuation, $1/G_{\text{src}}$ & 0.11 & \\
    Common coupled cavity build--up, $G_+$ & 5000 & \\
    Differential coupled cavity build--up, $G_-$ & 31.4 & \\
    Common coupled cavity pole, $f_+$ & 0.6 & Hz \\
    Differential coupled cavity pole, $f_-$ & 335--390 & Hz \\
    RF modulation index & 0.13--0.26 & rad\\
    Test mass diameter & 34 & cm \\
    Test mass thickness & 20 & cm \\
    Beam size at end test mass & 6.2 & cm \\
    Beam size at input test mass & 5.3 & cm \\
    Mass of the test mass, $M$ & 40 & kg \\
    \end{tabular}
    \end{ruledtabular}
\end{table}

Several critical improvements distinguish Advanced LIGO from the initial detectors~\cite{design_aligo_peter}. The much improved seismic isolation system ~\cite{isi_matichard} reduces the impact of ground vibrations. All photodetectors, used in the observing mode, are installed in vacuum to avoid the coupling of ambient acoustic noise to the gravitational wave channel. The larger and heavier test masses lead to a reduction of quantum radiation pressure induced motion and thermal noise~\cite{design_aligo}. Multi-stage pendulums with a monolithic lower suspension stage~\cite{sus_aston} filter ground motion and improve suspension thermal noise. Furthermore, instead of using coil-magnet actuation pairs to exert control forces on the test masses, electrostatic interaction is employed. This actuation scheme helps to avoid coupling of magnetic noise to the gravitational wave channel~\cite{bosem_garbone, esd_derosa}.

Lower arm cavity loss, coupled with an increase in the available power from the Nd:YAG laser, allows up to 800\,kW of laser power to circulate in the arm cavities---20~times higher than in initial LIGO $-$ significantly reducing the high frequency quantum noise. The use of optically stable folded recycling cavities allows for better confinement of the spatial eigenmodes of the optical cavities \cite{stable-cavities}. The signal recycling cavity~\cite{future_detune_src}, which was not present in initial LIGO, was introduced at the antisymmetric port to broaden the frequency response of the detector and improve its sensitivity at frequencies below 80\,Hz and above 200\,Hz.

Because O1 was the first observing run, and work remains to be done on the detectors to bring them to their design sensitivity, not all of the interferometer parameters were at their design values during O1. Most notably, the laser power resonating in the arm cavities was 100\,kW instead of the planned 800\,kW. More power in the arm cavities improves the shot noise level as discussed in Sec.~\ref{noises_quantum}. Circulating optical power will be increased in future observational runs. Additionally, the signal recycling mirror transmissivity was 36\%, in contrast to the design value of 20\%. This higher transmissivity of the signal recycling mirror improves the quantum noise in the frequency range from 60\,Hz to 600\,Hz at the price of reducing the sensitivity at other frequencies. Finally, the best measured Advanced LIGO sensitivity in the frequency range 20--100\,Hz, as discussed in Sec.~\ref{sec:instnoise}, is limited by a wide range of understood technical noise sources as well as currently unknown noise sources.

Fig.~\ref{fig:sensitivity} shows the Advanced LIGO detector's sensitivity during the first observing run. The performance of both the L1 and H1 detectors is compared to the initial LIGO sensitivity and the design sensitivity: the improvement with respect to S6 was 3--4 times at 100\,Hz and higher frequencies. Below 100\,Hz, the upgraded seismic isolation system yielded even larger improvements, with more than an order-of-magnitude-better strain sensitivity for frequencies below 60\,Hz. The sensitivity of Advanced LIGO can also be quantified as maximum distance at which a given astrophysical source would be detectable, known as ``horizon distance''. Fig.~\ref{fig:range} shows the horizon distance as a function of the chirp mass for coalescence of neutron star ($\mathcal{M} \lesssim 2 M_\odot$) and black hole ($\mathcal{M} \gtrsim 2 M_\odot$) binaries. For chirp masses $\lesssim 100 M_\odot$ horizon distance increases with chirp mass since gravitational wave signal is stronger from heavier binary systems. However, the signal also shifts towards lower frequencies (and out of LIGO frequency band) for massive binary systems, and horizon distance decreases for chirp masses $\gtrsim 100 M_\odot$.

\subsection{Interferometer Controls}

In operation the laser light needs to resonate inside the optical cavities. This requires that the residual longitudinal motion of the optical cavities be kept within a small fraction of the laser wavelength \cite{lock_aligo_als}. The suspended mirrors naturally move by $\sim 1 \mu \mbox{m}$ at the microseismic frequencies around 100\,mHz---much larger than the width of a resonance. To suppress this motion, a sophisticated length sensing and control system is employed, using both the well-known Pound-Drever-Hall technique \cite{design_pdh, design_schnupp} and a version of homodyne detection known as ``DC readout'' \cite{ifo_dc_readout}. Table \ref{table:requirements} shows linewidths and requirements for residual root-mean-square (RMS) motion of the main interferometric degrees of freedom.

\begin{table}[t]
    \caption{The linewidths of Pound-Drever-Hall signals and the requirements for residual RMS motion for the main interferometric degrees of freedom.}\label{table:requirements}
    \begin{ruledtabular}
    \begin{tabular}{l r r}
    Degree of freedom & Linewidth & Residual \\ \hline
    Common arm length & 6~pm & 1~fm \\
    Differential arm length & 300~pm & 10~fm\\
    Power recycling cavity length & 1~nm & 1~pm\\
    Michelson length & 8~nm & 3~pm\\
    Signal recycling cavity length & 30~nm & 10~pm\\
    \end{tabular}
    \end{ruledtabular}
\end{table}

An electro-optic modulator generates radio frequency (RF) phase modulation sidebands at 9\,MHz and 45\,MHz, symmetrically spaced about the laser carrier frequency. The Pound-Drever-Hall technique is used to sense all longitudinal degrees of freedom except for the differential arm channel. Feedback control signals actuate on the suspended mirrors, using either coil-magnet or electrostatic actuation. The common arm cavity length is also used as a reference to stabilize the laser frequency, with sub-mHz residual fluctuations (in detection band).

The gravitational wave signal is extracted at the anti-symmetric port of the interferometer, where fluctuations in the differential arm cavity length are sensed. The arm cavities are held slightly off-resonance by an amount referred to as the differential arm offset $\Delta L$. This offset of roughly 10\,pm generates the local oscillator field, which is necessary for the DC readout. An output mode cleaner \cite{koji_omc} located between the antisymmetric output and the homodyne readout detectors, is used to filter out the RF sidebands as well as any higher-order optical modes, as these components do not carry information about the differential arm cavity length.

A similar feedback control scheme is employed to keep the optical axes aligned relative to each other and the laser beam centered on the mirrors~\cite{ifo_asc}. This system is required to maximize the optical power in the resonant cavities and keep it stable during data collection. A set of optical wavefront sensors is used to sense internal misalignments~\cite{design_wfs_nergis}. At the same time, DC quadrant photodetectors sense beam positions relative to a global reference frame. The test mass angular motions are stabilized to 3\,nrad~rms, keeping power fluctuations in the arm cavities smaller than 1\% on the time scale of a few hours.

\subsection{Strain Calibration}

For the astrophysical analyses, the homodyne readout of the differential arm cavity length needs to be calibrated into dimensionless units of strain~\cite{cal_s5}. This is complicated by the fact that the feedback servo for this degree-of-freedom has a bandwidth of about 100\,Hz, extending well into the band of interest. Denoting the control signal sent to the end test masses with $s$, and the error signal, as measured by the photodiodes in units of W, with $e$, the strain signal $h$ is
\begin{equation}
	h = A s + C^{-1} e,
\end{equation}
where $A$ is the calibration of the actuator strength into strain and is computed using dynamical models. Transfer function $C$ is the optical response from strain to the error signal and is given by
\begin{equation}\label{eq:optical_gain}
	C(f) = \frac{4 \pi G_{\text{arm}} L_0}{\lambda} 
	\left( \frac{G_{\text{prc}} P_{\text{in}} P_{\text{LO}}} {G_{\text{src}}} \right)^{1/2} K_-(f),
\end{equation}
where $P_{\text{in}}$ is the interferometer input power and $P_{\text{LO}}$ is power of the local oscillator coming out from the interferometer. Signal recycling cavity gain $G_{\text{src}}=9.2$ is in the denominator since differential arm signal is anti resonant in this cavity.

\begin{figure}[tb]
    \centering
    \includegraphics[width=0.45\textwidth]{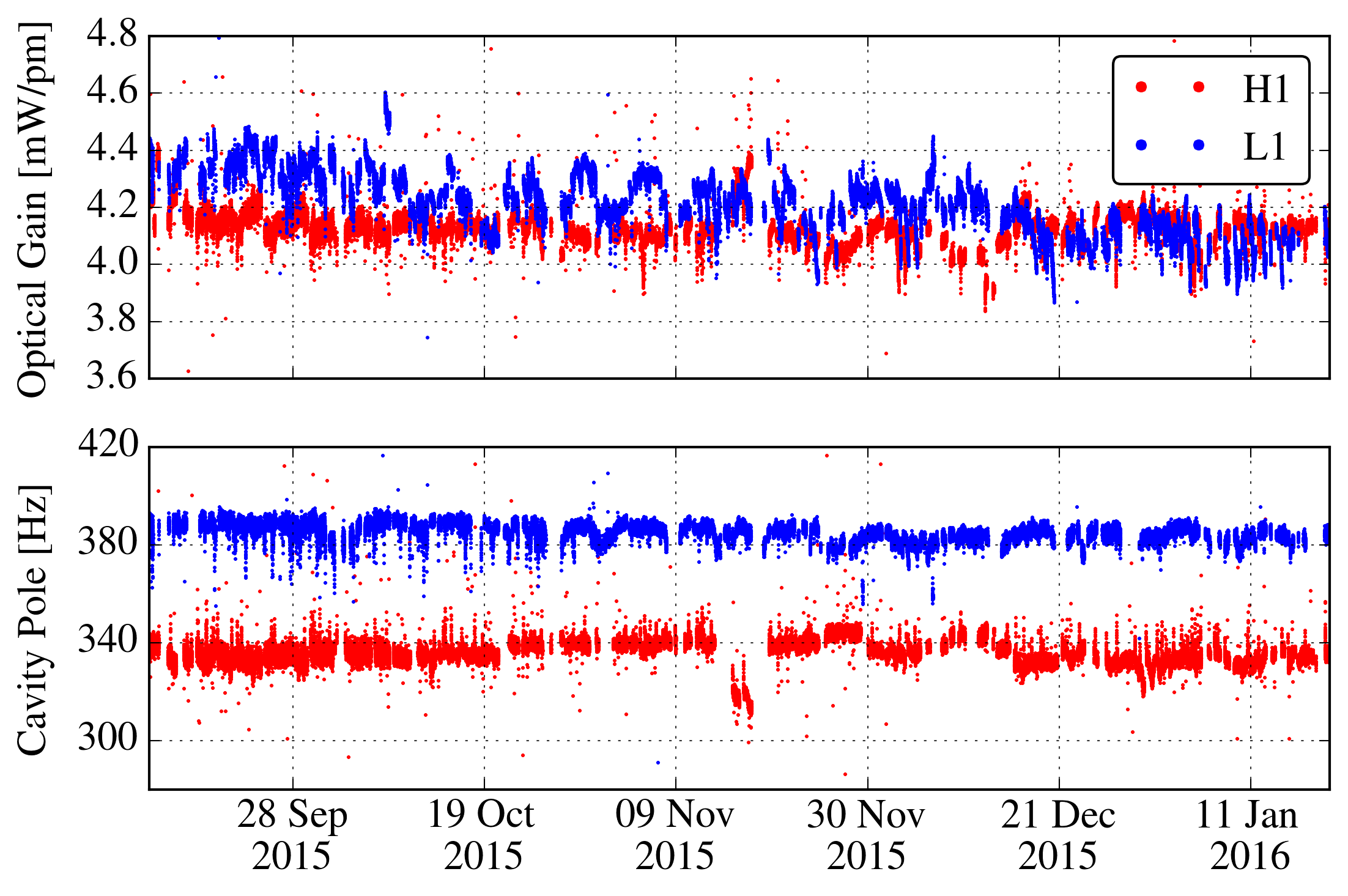}
    \caption {Time-varying response of the Advanced LIGO detectors. The top panel shows the optical gain variations over a time span of one month, whereas the bottom panel shows the variations of the differential coupled cavity pole frequency over the same time span. The blue traces are for the LIGO Livingston Observatory (L1) and the red traces for the LIGO Hanford Observatory (H1).}
    \label{fig:data_kappac}
\end{figure}

Ideally, the actuator transfer function $A$ is stable over time. In practice, a time-varying charge accumulates on the test masses, changing the actuation strength and introducing noise into the gravitational wave channel (see Sec.~\ref{charging_noise}). The optical transfer function $C$ is also non-stationary, being modulated mainly by angular motion of the test masses.

The optical response $C$ is tracked using a system known as the ``photon calibrator'', which consists of an auxiliary $\mathrm{Nd}^{3+}\mathrm{:YLF}$ laser (operating at a wavelength of 1047~nm), an acousto-optic modulator, and a set of integrating spheres~\cite{cal_pcal}. This calibration system actuates on the end test masses, applying set of sinusoidal excitations via radiation pressure, to track variations of the optical gain and of the differential coupled cavity pole frequency. Three weeks of such data are shown in Fig.~\ref{fig:data_kappac}, showing that the optical response of the detectors is stable over time.

The absolute accuracy of the photon calibrator is limited by the uncertainties in its photodetector calibration, as well as any optical losses between the test mass and the photodetector. Overall, the uncertainty in the calibration of the interferometer over the entire operational frequency range from 10\,Hz$-$5\,kHz is estimated to be smaller than 10\% and 10 degrees~\cite{ligo_cal_2016}.

\section{Analysis of the Instrumental Noise}
\label{sec:instnoise}

\begin{figure}[t!]
\centering
    \begin{minipage}{0.47\textwidth}
        \includegraphics[width=\textwidth]{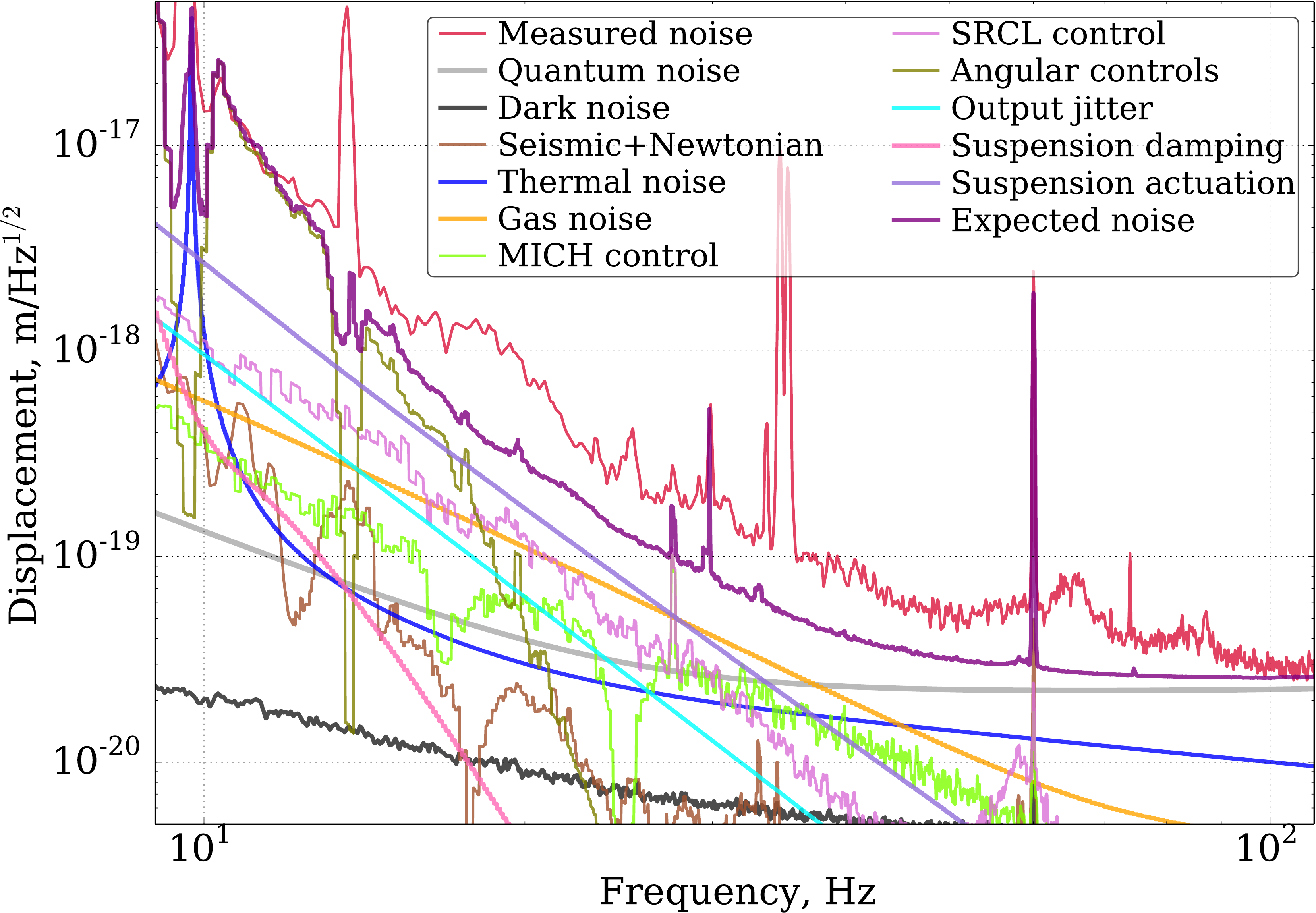}
    \end{minipage}
    \begin{minipage}{0.47\textwidth}
        \vspace{0.15cm} \mbox{(a) LIGO Livingston Observatory}
    \end{minipage}
    \begin{minipage}{0.47\textwidth}
        \vspace{0.2cm}\includegraphics[width=\textwidth]{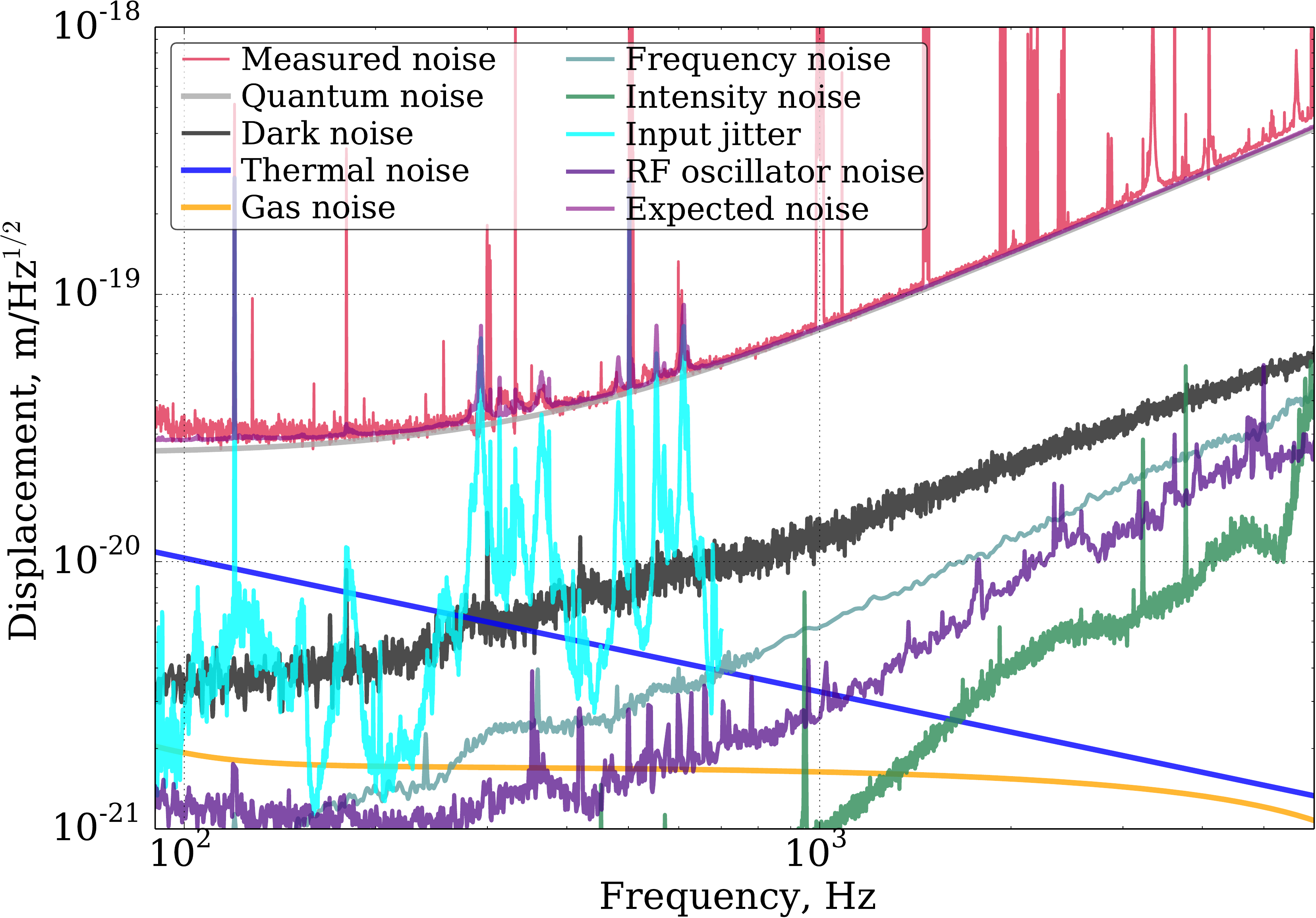}
    \end{minipage}
    \begin{minipage}{0.47\textwidth}
        \vspace{0.15cm} \mbox{(b) LIGO Hanford Observatory}
    \end{minipage}
    \caption[Noise Budget.]{Noise budget plots for the gravitational wave channels of the two LIGO detectors. The strain sensitivities are similar between the two sites. Plot~(a) shows the low-frequency curves for L1, whereas Plot~(b) shows the high-frequency curves for H1 detector. Quantum noise is the sum of the quantum radiation pressure noise and shot noise. Dark noise refers to electronic noise in the signal chain with no light incident on the readout photodetectors. Thermal noise is the sum of suspension and coating thermal noises. Gas noise is the sum of squeezed film damping and beam tube gas phase noises. The coupling of the residual motion of the Michelson (MICH) and signal recycling cavity (SRCL) degrees of freedom to gravitational wave channel is reduced by a feedforward cancellation technique. At low frequencies, there is currently a significant gap between the measured strain noise and the root-square sum of investigated noises. At high frequencies, the sensitivity is limited by shot noise and input beam jitter.}
    \label{fig:data_nb}
\end{figure}

The calibrated gravitational wave signal is compared to the known noises in order to understand what limits the sensitivity of the instrument as a function of frequency. Fig. \ref{fig:data_nb} summarizes the noise contributions from various sources to the gravitational wave channel for the Livingston and Hanford detectors. The coupling of each noise source to the gravitational wave channel at a frequency $f$ is estimated using the following equation:
\begin{equation}
	L(f) = L_0 h(f) = T(f) \times N(f),
\end{equation}
where $N(f)$ is the noise spectrum measured by an auxiliary (witness) sensor or computed using analytical model, and $T(f)$ is the measured or simulated transfer function from this sensor to the gravitational wave channel.

Noise sources can be divided into classes according to their origins and coupling mechanisms~\cite{rana_thesis, data_noise}. One clear way to differentiate noises is to split them into displacement and sensing noises: displacement noises cause real motion of the test masses or their surfaces, while sensing noises limit the ability of the instrument to measure test mass motion. However, this distinction is not perfect, since some noise sources (e.g.,~laser amplitude noise) can be assigned to both categories, as discussed in Sec.~\ref{noises_laser}.

Another way to classify noise sources is to divide them into fundamental, technical and environmental noises. Fundamental noises can be computed from first principles, and they determine the ultimate design sensitivity of the instrument. This class of noises, which includes thermal and quantum noise, cannot be reduced without a major instrument upgrade, such as the installation of a new laser or the fabrication of better optical coatings. Technical noises, on the other hand, arise from electronics, control loops, charging noise and other effects that can be reduced once identified and carefully studied. Environmental noises include seismic motion, acoustic and magnetic noises. The design of Advanced LIGO calls for the contributions of technical and environmental noises to the gravitational wave channel to be small compared to fundamental noises. In practice, the sensitivity can be reduced due to unexpected noise couplings. Many technical and environmental noises have been identified and are discussed in the following sections. At the same time, the dominant noise contributor in the frequency range 20--100\,Hz has not yet been identified.

The narrowband features in the sensitivity plots shown in Fig.~\ref{fig:data_nb} are caused by power lines (60\,Hz and harmonics), suspension mechanical resonances, and excitations that are deliberately added to the instrument for calibration and alignment purposes. These very narrow lines are easily excluded from the data analysis, while the broadband noise inevitably limits the instrument sensitivity. The latter is therefore a more important topic of investigation.

\subsection{Seismic and thermal noises}
\label{seismic_thermal}

Below 10\,Hz, there is significant displacement noise from residual seismic motion. On average, at both the Livingston and Hanford sites, the ground moves by $\sim 10^{-9}~\text{m}/\sqrt{\text{Hz}}$ at 10\,Hz---ten orders of magnitude larger than the Advanced LIGO target sensitivity at this frequency. To address this difference, seismic noise is filtered using a combination of passive and active stages. The test masses are suspended from 
quadruple pendulums \cite{sus_aston}. These passive filters have resonances as low as 
0.4\,Hz and provide isolation as $1/f^8$ in the detection bandwidth.  The 
pendulums are mounted on multistage active platforms \cite{isi_part_1, isi_part_2}. These 
systems use very-low-noise inertial sensors to provide the required isolation in the detection band and at lower frequencies (below 10\,Hz). This isolation is crucial for bringing the interferometer into the linear regime and allowing the longitudinal control system to maintain it on resonance. The active platforms combine feedback and feedforward 
control to provide one order of magnitude of isolation at the microseism 
frequencies (around  0.1 Hz) and three orders of magnitude between 1\,Hz 
and 10\,Hz. Most of the suspension resonances 
are located in this band, where ground excitation from anthropogenic noise and wind is significant.

Fluctuations of local gravity fields around the test masses---caused by ground motion and vibrations of the buildings, chambers, and concrete floor---also couple to the gravitational wave channel as force noise \cite{data_newtonian} (gravity gradient noise). The coupling to the differential arm length displacement is given by
\begin{equation}
\begin{split}
	& L(f) = 2 \frac{N_{\text{grav}}(f)}{(2\pi f)^2} \\
	& N_{\text{grav}}(f) = \beta G \rho N_{\text{sei}}(f),
\end{split}
\end{equation}
where $N_{\text{grav}}$ is the fluctuation of the local gravity field projected on the arm cavity axis, the factor of 2 accounts for the incoherent sum of noises from the four test masses, $G$ is the gravitational constant, $\rho \simeq 1800 ~\mathrm{kg ~ m^{-3}}$ is the ground density near the mirror, $\beta \simeq 10$ is a geometric factor, and $N_{sei}$ is the seismic motion near the test mass. Since the ground near the test masses moves by $\simeq 10^{-9} \mathrm{m/\sqrt{Hz}}$ at 10\,Hz,
 local gravity fluctuations at this frequency are $N_{\text{grav}} \approx 10^{-15} \mathrm{m ~ s^{-2} /\sqrt{Hz}}$ and the total noise coupled into the gravitational wave channel at 10\,Hz is $L \approx 5 \times 10^{-19}$\,$\mathrm{m/\sqrt{Hz}}$. 
Gravity gradient noise is one of the limiting noise sources of the Advanced LIGO design in the frequency range 10--20\,Hz. However, the typical sensitivity measured during O1 is still far from this limitation.

Thermal noises arise from finite losses present in mechanical systems and couple to the gravitational wave channel as displacement noises. Several sources of thermal noise can be identified. Suspension thermal noise~\cite{data_thermal_thesis} causes motion of the test masses due to thermal vibrations of the suspension fibers. Coating Brownian noise is caused by thermal fluctuations of the optical coatings, multilayers of silica and titania-doped tantala~\cite{data_thermal_coating_loss, data_thermal_coating, data_thermal_coating_equation, harry_coating_2012}. Thickness of the coatings was optimized to reduce their thermal noise and provide the required high reflectivity of the mirrors~\cite{agresti_coating_2006, villar_coating_2010}. Thermal noise also arises in the substrates of the test masses~\cite{data_thermal_sub_brag, data_thermal_sub_liu}, but this effect is less significant. Thermal noise levels are analytically computed using the fluctuation-dissipation theorem \cite{data_thermal_levin} and independent measurements of the losses of materials. The model predicts that thermal noise limits the Advanced LIGO design sensitivity in the frequency band 10--500\,Hz, but is below current sensitivity by a factor of $\geq 3$.

\subsection{Quantum noise}
\label{noises_quantum}

Quantum noise is driven by fluctuations of the optical vacuum field entering the interferometer through the antisymmetric port~\cite{PhysRevLett.45.75, design_caves}. This
fundamental noise couples to the interferometer sensitivity in two complementary ways~\cite{Braginsky_1992}.
For one, vacuum fluctuations disturb the
optical fields resonating in the arm cavities, creating displacement noise by exerting a fluctuating radiation
pressure force that physically moves the test masses~\cite{data_quantum_sr,
  data_thesis_corbitt}. The vacuum field is amplified by the optical cavities, and the noise seen in the differential arm channel is given by:
\begin{equation}
\begin{split}
	L(f) = & \frac{2}{c M \pi^2 f^2} \left( h \nu G_- P_{\text{arm}} \right)^{1/2} K_-(f) \\
	L(f) = & \frac{1.38 \times 10^{-17}}{f^2} \left( \frac{P_{\text{arm}}}{100\text{\,kW}} \right)^{1/2} K_-(f) \frac{\text{m}}{\sqrt{\text{Hz}}},
\end{split}
\end{equation}
where $h$ is Planck constant and $P_{\text{arm}}$ is the power circulating in the arm cavities. This ``quantum radiation pressure noise'' imposes a fundamental limit to the design sensitivity below 40\,Hz, though it is still far from being a concern at the present operating power~\cite{design_aligo}.

The vacuum fluctuations entering interferometer through the antisymmetric port also introduce shot noise in the gravitational wave channel~\cite{lisa_squeezed}. Vacuum fluctuations also mix with the main beam due to optical losses between the interferometer and the photodetector. In the current state of Advanced LIGO 25\% of power at the antisymmetric port is lost due to the output Faraday isolator, mode mis-match of the beam into the output mode cleaner cavity, and imperfect quantum efficiencies of the photodetectors. So the fraction of the power that is transmitted to the photodiodes is $\eta=0.75$.

Differential arm sensing noise due to shot noise on the photodetectors can be written as $L(f) = L_0 N_{\text{shot}} / C(f)\eta$, where  $N_{shot} = (2 h \nu \eta P_{\text{LO}})^{1/2}$ is the shot noise on the photodetector in units of W/$\sqrt{\text{Hz}}$. The signal transfer function $C(f)$ is determined by Eq.~\ref{eq:optical_gain}. The total shot noise is given by equations
\begin{equation}
\begin{split}
	L(f) & = \frac{\lambda}{4 \pi G_{\text{arm}}} 
		\left( \frac{2 h \nu G_{\text{src}}}{G_{\text{prc}} P_{\text{in}} \eta} \right)^{1/2} 
		\frac{1}{K_-(f)} \\
        L(f) & = 2 \times 10^{-20} \left( \frac{\mathrm{100\,kW}}{P_{\text{arm}}
          \eta} \right)^{1/2}
	\frac{1}{K_-(f)}
        \frac{\mathrm{m}}{\sqrt{\mathrm{Hz}}}.
\end{split}
\end{equation}
Local oscillator power $P_{\text{LO}}$ cancels out in the final equation, and shot noise level is independent of the differential arm offset for small offsets $\Delta L \lesssim 100$\,pm.

The Advanced LIGO optical configuration is tuned to maximize power circulating in the arm cavities. Common coupled cavity build--up (ratio between the power resonating in the arms and power entering the interferometer) is related to the losses in the arm cavities by
\begin{equation}
	G_{\text{comm}} \lesssim \frac{1}{2 Y_{\text{arm}}},
\end{equation}
where $Y_{\text{arm}}$ is round trip optical loss in one arm. During O1 the power circulating in the arm cavities was $G_{\text{comm}}\simeq 5000$ greater than the power entering the interferometer, corresponding to a round trip optical loss of $Y_{\text{arm}} \simeq 100$\,ppm in each arm cavity.  The target optical gain for Advanced LIGO was 7500, which corresponds to round trip losses in the arm cavities of about 75\,ppm. This number can possibly be achieved once the test masses are replaced after the second science run. The discrepancy in the round trip losses between the predicted and measured values is currently under study. Shot noise limits the design sensitivity above 40\,Hz, and the current sensitivity above 100\,Hz.

\subsection{Gas noise}

The Advanced LIGO optics are located inside vacuum chambers. The gas pressure in the corner station, where the dual-recycled Michelson interferometer is housed, and in the 4-km arm tubes, is maintained below $10^{-6}$\,Pa. The presence of residual gas causes both displacement and sensing noise: thermal motion of gas molecules inside the vacuum chambers results in momentum exchange with the test masses via collisions; meanwhile, forward scattering of photons by the gas molecules in the arm tubes modulates the optical phase of the beam. 

\subsubsection{Squeezed film damping}

Residual gas in the vacuum system exerts a damping force on the test masses and introduces displacement noise~\cite{data_gas_film}.
This noise is amplified by a factor of $\sim$10 below 100\,Hz due to the small gap of 5\,mm between the end test and reaction masses \cite{data_gas_unconstrained} (the top view of a test mass and its surroundings is shown in Fig. \ref{fig:data_charge_etm}). The total noise can be estimated by applying the fluctuation-dissipation theorem or by running a Monte Carlo simulation \cite{data_gas_matt}. 
The coupling coefficient depends on the gas pressure and the molecular mass, and it is found to be (below 100\,Hz)
\begin{equation}
\begin{split}
	& F(f) = 1.5 \times 10^{-14} \left( \frac{p}{10^{-6}} \right)^{1/2}
	\left( \frac{m}{m_{H_2}} \right)^{1/4} \frac{\text{N}}{\sqrt{\text{Hz}}},
\end{split}
\end{equation}
where $p$ is the residual gas pressure in Pa, and $m$ is the mass of a gas molecule.
The calculated squeezed film damping noise shown in Fig.~\ref{fig:data_nb} (a)
is the sum of contributions from nitrogen ($p_{N_2} \approx 6\times 10^{-7}$\,Pa), hydrogen ($p_{H_2} \approx 2\times 10^{-6}$\,Pa) and water ($p_{H_2 O} \approx 10^{-7}$\,Pa). 

\begin{figure}[ht]
	\centering
	\includegraphics[width=8.5cm]{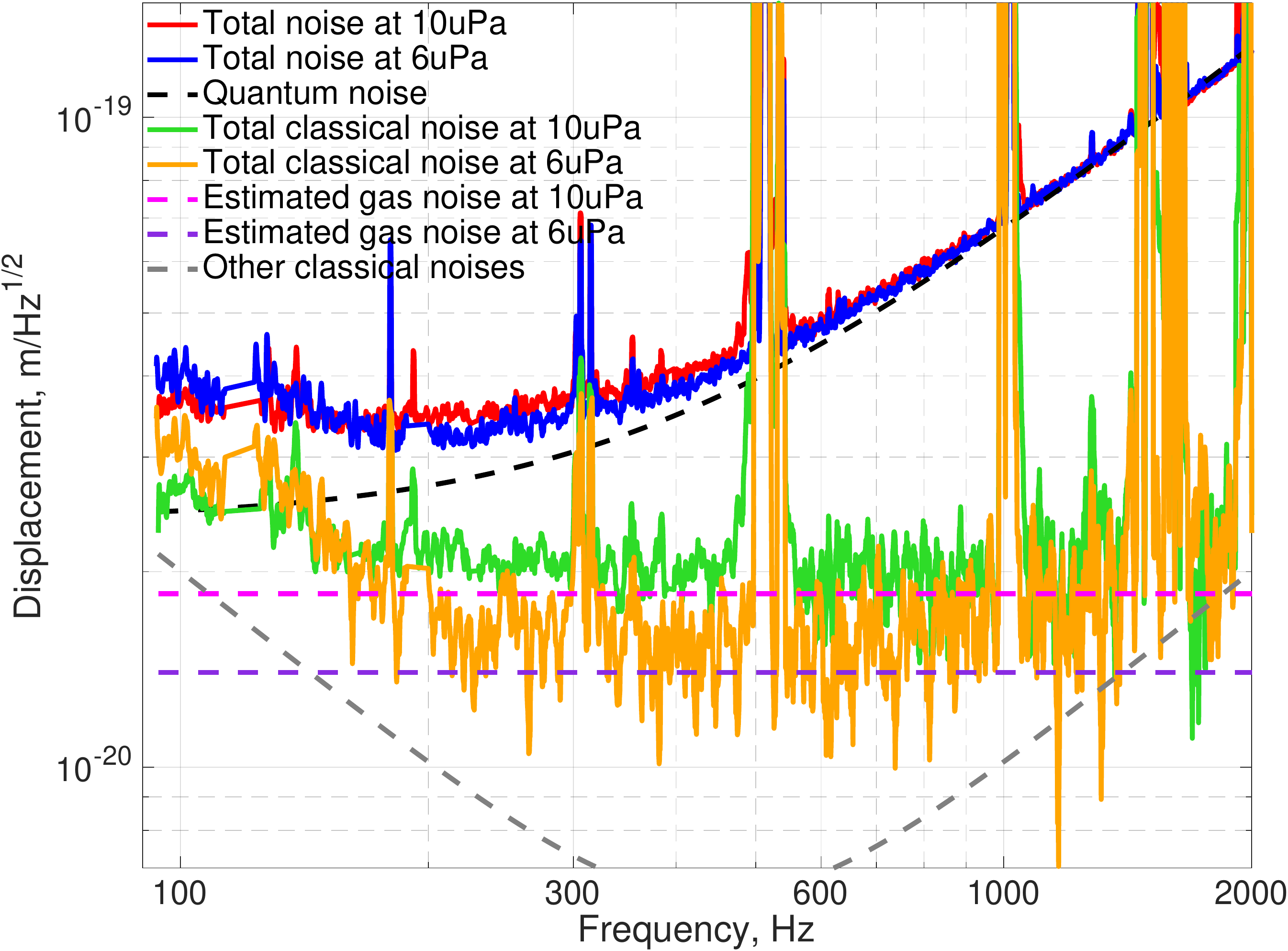}
	\caption {Measurement of the gas phase noise for the two different pressures (average values are 10\,uPa and 6\,uPa) in one 4\,km arm of the Livingston detector. The red and blue traces show total measured noise before and after the pump down. The dashed black curve shows the quantum noise level, which is independent of the pressure in the arms. The green and orange curves show total classical noise at pressure 10\,uPa and 6\,uPa correspondingly. The magenta and violet curves show the estimated gas phase noises. Reduction of classical noise is in agreement with the model that was used to compute gas phase noise. The gray curve shows other classical noises which do not depend on the gas pressure. Below 300\,Hz there is an unknown 1/f noise. At higher frequencies, classical noise grows with frequency, and is dominated by dark noise of the photodetectors and laser frequency noise.}
	\label{fig:data_vac}
\end{figure}

\subsubsection{Phase noise}

Phase noise induced by the stochastic transit of molecules through the laser beam in the arm cavities, can be modeled by calculating the impulsive disturbance to the phase of the laser field as a gas molecule moves through the beam \cite{data_gas_phase}. Such a model was used to estimate the high frequency part of gas noise curve shown in Fig.~\ref{fig:data_nb} (b).
This estimation accounts for the pressure distribution in the arm cavities along with the profile of the laser beam, with the most significant noise contribution coming from the geometrical center of the tube, where the beam waist is located.
The expected noise from residual gas is given by 
\begin{equation}
\begin{split}
	& L(f) = 4 \times 10^{-21}  N_{\text{gas}} \frac{\text{m}}{\sqrt{\text{Hz}}} \\
	& N_{\text{gas}} = \left( \frac{a_{\text{gas}}}{a_{H_2}} \right) \left( \frac{m_{\text{gas}}}{m_{H_2}} \right) ^{1/4}
	\left( \frac{p}{10^{-8}} \right)^{1/2},
\end{split}
\end{equation}
where $a_{\text{gas}}$ is the polarization of the gas molecules.

The estimation of the gas phase noise was verified by changing the pressure in one of the arms by a factor of 3 at the end station and factor of 1.7 at the half-way point. A variation of differential arm noise was measured using relative intensity fluctuations at the output port, as shown in Fig. \ref{fig:data_vac}. Though, as discussed in Sec.~\ref{noises_quantum}, the sensitivity above 100\,Hz is limited by shot noise, classical noise can be revealed by incoherent subtraction of shot noise from the measured signal. Using this technique, classical noise was observed to change during this test as predicted by the model.

\subsection{Charging noise}
\label{charging_noise}

\begin{figure}[ht]
	\centering
	\includegraphics[width=8cm]{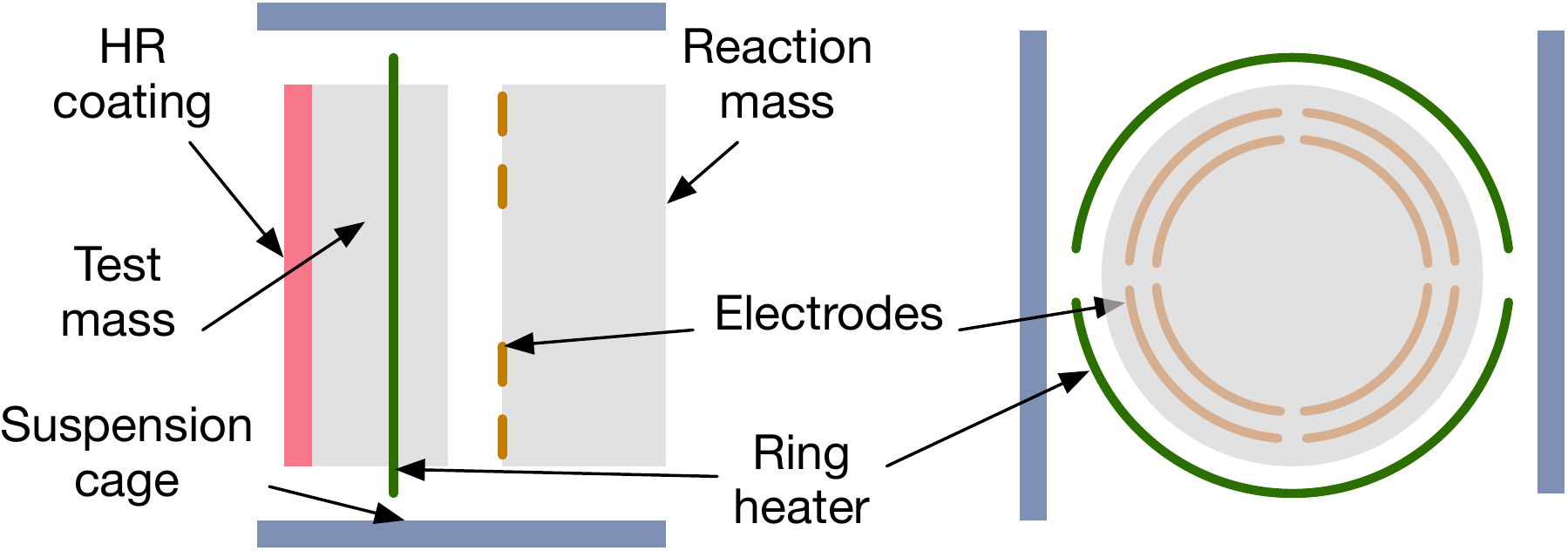}
	\caption{Top and front views of a test mass showing the arrangement of the electrodes, high reflective (HR) coating, ring heater and surrounding metal cage. Electrodes are used for actuation on the test mass. The ring heater is used to correct the curvature of the mirror.}
	\label{fig:data_charge_etm}
\end{figure}

During the Advanced LIGO commissioning, it was discovered that the electrostatic actuation on the test masses was not symmetric among the four electrodes located on the reaction mass (see Fig. \ref{fig:data_charge_etm}). This mismatch in actuation strength is caused by electrostatic charge~\cite{geo_charge}, which is distributed on the test masses in a non-uniform manner and is time dependent.

Ideally, there should be no charge on the test masses, except for the one accumulated due to electrostatic actuation. However, some electric charge may be left by imperfect removal of the First Contact polymer used for cleaning and protection of the optics~\cite{first_contact}. Moreover, surfaces of the test masses also lose electrons due to UV photons, generated by nearby ion pumps used in the vacuum system. Dust particles in the vacuum system provide yet another source of charging. It was discovered that the charge distribution changes on the week time scale. An order-of-magnitude estimate of the charge density on the front and back surfaces of the end test masses is $\sigma \sim 10^{-11}$ C/$\rm{cm}^2$. This number was achieved by exciting the electrodes and the potential of the ring heaters while measuring the longitudinal and angular motion of the test mass. 

There are two coupling mechanisms of charging noise to the gravitational wave channel. The first mechanism arises due to interaction of the time -- variant charge with the metal cage around the test mass. The second coupling mechanism comes from voltage
fluctuations of the various pieces of grounded metal in the vicinity of the test mass. Voltage noise creates fluctuations of the electric field $E$ and applies a force $F_{ch}$ on the test mass according to the following equation
\begin{equation}\label{eq:charge}
	F_{ch} = \int E \sigma dS,
\end{equation}
where the integral is computed over both the front and back surfaces of the test mass. In this paper, we consider only the second coupling mechanism, since it is estimated to be the dominant one.

The broadband voltage noise on the ground plane is measured to be roughly 1\,uV/$\sqrt{\text{Hz}}$. This number was measured between the grounded suspension cage and the floating ring heaters. Since the characteristic distance between the test masses and the metal cage is 10\,cm, the fluctuations in the electric field near the test mass are $\sim 10^{-5} \text{\,V/m/}\sqrt{\text{Hz}}$. The total noise coupling above 10\,Hz is estimated using the equation
\begin{equation}
	L(f) = \frac{F_{ch}}{M (2 \pi f)^2} \approx \frac{10^{-16}}{f^2} 
	\frac{\sigma}{10^{-11} \text{C}/\text{cm}^2} \frac{\text{m}}{\sqrt{\text{Hz}}}.
\end{equation}

The coupling of voltage fluctuations on the ground plane to the gravitational wave signal was reduced by a factor of 10--100 by discharging the test masses. Charge from the front surface can be efficiently removed using ion guns \cite{data_discharge_ugolini, data_discharge}: positive and negative ions are introduced into the chamber, when the pressure inside is $\sim 10^3$\,Pa, and annihilate surface charges on the front surface of the test mass. During the discharge procedure, it was found that the ions cannot efficiently reach the back surface due to the small gap between the test mass and the reaction mass, as shown in Fig. \ref{fig:data_charge_etm}. The back surface of the end test masses was discharged by opening the chambers, separating the test and reaction mass, and directing an ion gun at close range towards the surfaces in the gap.

\subsection{Laser amplitude and frequency noise}
\label{noises_laser}

Advanced LIGO employs a Nd:YAG nonplanar ring oscillator as the main laser~\cite{design_psl}. Intensity and frequency fluctuations of such a laser can be roughly approximated as $10^{-4}/f~/\sqrt{\text{Hz}}$ and $10^4/f\text{\,Hz}/\sqrt{\text{Hz}}$, respectively, in the frequency range 10\,Hz$-$5\,kHz. In the same band, the Advanced LIGO requirements are $\sim 10^{-8}~/\sqrt{\text{Hz}}$ for intensity noise and $\sim 10^{-6}\text{\,Hz}/\sqrt{\text{Hz}}$ for frequency noise. 
In order to meet those requirements, a hierarchical control system is implemented. First of all, laser noises are actively suppressed using intensity and frequency stabilization servos. Additionally, laser noise on the beam entering the main interferometer is passively filtered by $K_+$ (Eq.~\ref{eq:poles}) due to the common-mode coupled cavity pole.

For laser amplitude noise, there are several coupling mechanisms.
First of all, the presence of the nonzero differential arm offset $\triangle L$ needed for the homodyne readout means that the carrier light at the antisymmetric port is directly modulated by amplitude noise entering the interferometer.
In addition, mismatches in the circulating arm powers and in the mirror masses also lead to intensity noise coupling through radiation pressure force at low frequencies (below 50\,Hz).

\begin{figure}[ht]
	\centering
	\includegraphics[width=8cm]{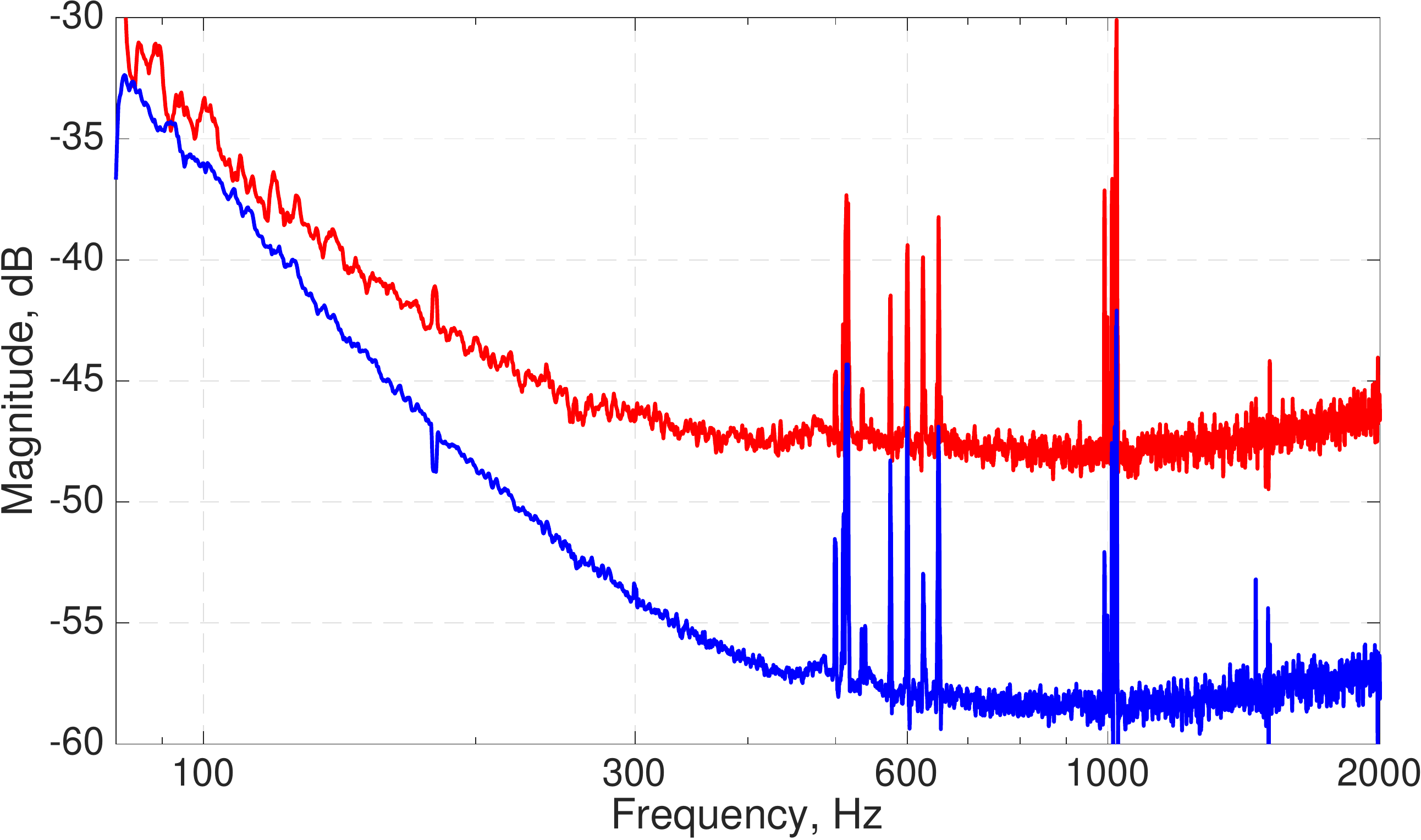}
	\caption{Measured transfer function of intensity fluctuations from interferometer input to the antisymmetric port.
        The blue trace corresponds to the case when substrate lenses of input test masses are matched.
        The red trace shows the coupling when substrate lenses are different by 7.5\,$\mu$D. For the difference of 40\,$\mu$D the coupling above 60\,Hz increases up to -25\,dB.}
	\label{fig:data_rin}
\end{figure}

Above 100\,Hz, the most significant broadband coupling of laser amplitude noise comes from unequal effective lenses in the input test masses, due to substrate inhomogeneity.
The presence of imbalanced lenses creates a direct conversion of the fundamental laser mode into higher-order spatial modes. As these modes do not resonate in the arm cavities, they are not filtered by the common-mode coupled cavity, and they therefore contribute to the coupling of laser intensity noise with a flat transfer function. 
A thermal compensation system (TCS)~\cite{aidan_tcs}, which employs auxiliary $\text{CO}_2$ laser beams and ring-shaped heating elements, has been installed to compensate for such imbalances. Fig.~\ref{fig:data_rin} shows that the coupling of intensity noise can be significantly reduced by equalizing the substrate lenses using the TCS system: if no correction is applied, the differential lens power is 40\,$\mu$D and the coupling coefficient at 300\,Hz is more than 40\,dB larger than the lowest value attainable with a proper TCS correction.

Laser frequency noise is largely cancelled at the antisymmetric port by virtue of the Michelson interferometer common-mode rejection ($\sim$ 1000 at 100\,Hz). However, residual frequency noise couples into the gravitational wave channel through the intentional asymmetry that is introduced into the Michelson interferometer to produce the necessary interference conditions for the RF control sidebands, and through imbalances in arm cavity reflectivities and pole frequencies~\cite{Sigg1997, Izumi2015b}.
The achieved laser frequency noise performance is limited primarily by sensing noises (shot noise, photodiode noise, and electronics noise) in the feedback control that stabilizes the laser frequency to the interferometer's common (mean) arm length. In Advanced LIGO, noise in the frequency stabilization error signals limits the residual frequency noise of the beam entering the main interferometer to $\simeq 10^{-6}~\text{Hz}/\sqrt{\text{Hz}}$ between 10 and 100\,Hz, and increasing as $f$ above 100\,Hz.

\subsection{Auxiliary Degrees-of-Freedom}

\begin{figure*}[t!]
\begin{center}
    \begin{minipage}{0.495\textwidth}
        \includegraphics[width = 0.99\textwidth]{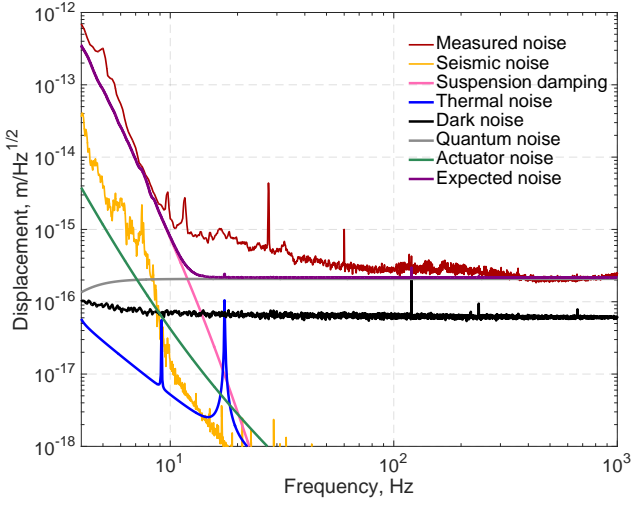}
    \end{minipage}
    \begin{minipage}{0.495\textwidth}
        \includegraphics[width = 0.99\textwidth]{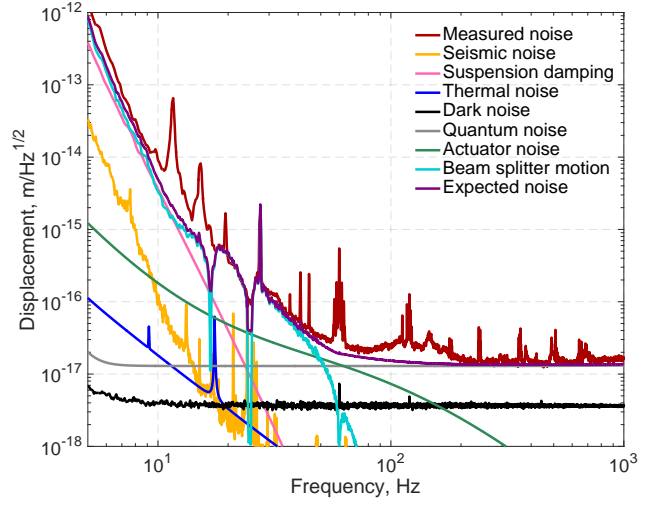}
    \end{minipage}
    \\ \vspace{2mm}
    \begin{minipage}{0.495\textwidth}
        \mbox{(a) Michelson length.}
    \end{minipage}
    \begin{minipage}{0.495\textwidth}
        \mbox{(b) Power recycling cavity length.}
    \end{minipage}
    \\ \vspace{5mm}
    \begin{minipage}{0.495\textwidth}
        \includegraphics[width = 0.99\textwidth]{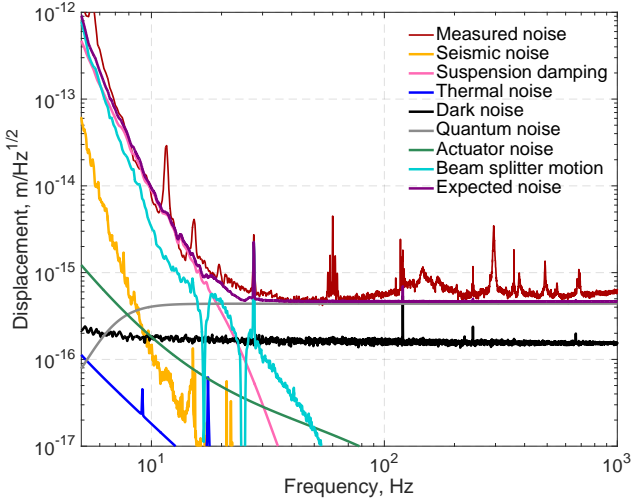}
    \end{minipage}
    \begin{minipage}{0.495\textwidth}
        \includegraphics[width = 0.99\textwidth]{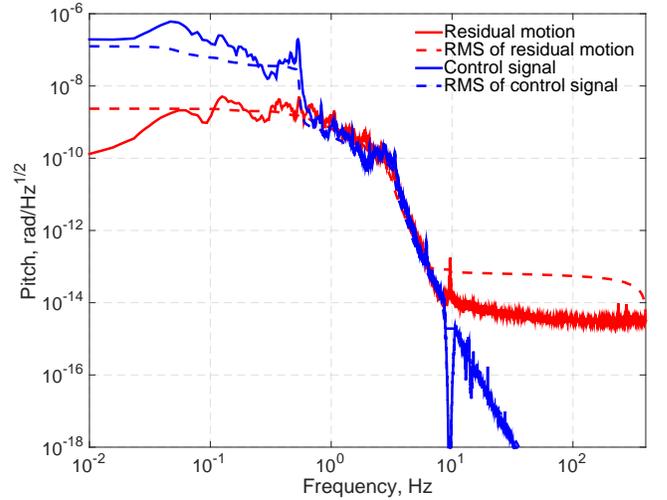}
    \end{minipage}
    \\ \vspace{2mm}
    \begin{minipage}{0.495\textwidth}
        \mbox{(c) Signal recycling cavity length.}
    \end{minipage}
    \begin{minipage}{0.495\textwidth}
        \mbox{(d) Test mass angular motion (pitch).}
    \end{minipage}

    \caption[Noise Budget for Auxiliary Degrees-of-Freedom]{Noise budgets for auxiliary degrees of freedom. Plot~(a) shows the noise curves for the Michelson length, Plot~(b) for the power recycling cavity length, Plot~(c) for the signal recycling cavity length, and Plot~(d) for the angular motion of one of the test masses in pitch. The signals are measured with the full interferometer operating in the linear regime. The most significant noise sources in the dual-recycled Michelson degrees of freedom are seismic noise, shot noise and electronics noise in the interferometric readout chains and in local sensors on the individual suspensions. Quantum noise in the signal recycling cavity length is signiticantly affected by the differential arm offset below 10\,Hz. In addition to coupling to the gravitational wave channel, auxiliary degrees of freedom also couple to each other. For example, beam splitter motion above 10\,Hz is caused by the Michelson control loop and dominates the power and signal recycling cavity length fluctuations in the frequency range 10-50\,Hz.}
    \label{fig:data_aux}
\end{center}
\end{figure*}

The use of a dual-recycled Michelson interferometer optimizes the detector response to gravitational waves. Additionally, active control of the mirror angular degrees of freedom is important to stabilize the interferometer optical response. However, any noise in the associated auxiliary degrees of freedom will couple to the gravitational wave channel at some level. Fig.~\ref{fig:data_aux} shows the typical noise in the auxiliary longitudinal degrees of freedom calibrated into displacement, as well as the typical angular noise in one of the arm cavity pitch degrees of freedom.

Any residual fluctuation of the Michelson length $N_{\text{mich}}$ couples to the transmitted power of the output mode cleaner, where the gravitational wave channel is transduced. The coupling mechanism is similar to that of a differential arm length fluctuation, but without the amplification factor provided by the arm cavity build-up $G_{\text{arm}}=270$:

\begin{equation}
	L(f) = \frac{1}{G_{\text{arm}}} N_{\text{mich}}(f).
\end{equation}
This coupling coefficient depends only weakly on the differential arm offset and alignment, unless the power build-up in the arm cavities is significantly changed.

Residual fluctuations of the signal recycling cavity length also couple to the gravitational wave channel, due to the differential arm offset $\Delta L$, through a radiation pressure force exerted on the test masses by the resonating optical fields. In the frequency range from 10 to 70\,Hz, the differential arm noise $L(f)$ due to signal recycling cavity longitudinal noise $N_{\text{srcl}}$ can be modeled as
\begin{equation}
	L(f) = \frac{0.16}{f^2} \frac{\Delta L}{10\text{\,pm}} N_{\text{srcl}}(f),
\end{equation}
where the numerical factor is determined mainly by the signal recycling mirror reflectivity and the masses of the cavity mirrors. Besides this linear coupling, a non-linear component appears due to low-frequency modulation of the differential arm offset $\Delta L$ (by $\simeq 10-20\%$), which arises from unsuppressed angular motion of the interferometer mirrors. Such motion generates higher-order mode content in the beam exiting the interferometer through the antisymmeteric port, leading to modulation of the power transmitted by the output mode cleaner and forcing the differential arm length servo to compensate by changing the offset $\Delta L$. At higher frequencies (above 70\,Hz), the coupling of the signal recycling cavity longitudinal noise depends on the mode matching between the signal recycling cavity and the arm cavities. This can be tuned using the thermal compensation system discussed above.

The coupling of the power recycling cavity length to the differential arm channel is caused by imbalances in the two arm cavities and cross couplings with other longitudinal degrees of freedom. Residual power recycling cavity length noise is less significant (by a factor of $\geq 10$) compared to other degrees of freedom of the dual—-recycled Michelson interferometer.

Finally, any residual angular motion of the test masses $N_{ang}$ couples to the gravitational wave channel geometrically due to beam mis-centering $d$ on the mirrors, according to the equation
\begin{equation}
	L(f) = d \times N_{\text{ang}}(f).
\end{equation}
The beam mis-centering itself is also modulated by the mirror angular motion $d = \bar{d} + d_{\text{ac}}$, where $\bar{d}$ and $d_{\text{ac}} \propto N_{\text{ang}}$ are stationary and non-stationary components of the beam position. For this reason, the coupling of the angular motion can be linear and non-linear. The angular feedback servos are optimized to suppress low-frequency motion of the cavity axis and $d_{\text{ac}}$ while avoiding injection of sensor noise at high frequencies.

The linear coupling of the auxiliary degrees of freedom to the gravitational wave channel is mitigated using a realtime feed-forward cancellation technique. Witness signals are properly reshaped using time-domain filters, and the cancellation signals are applied directly to the test masses. This feed-forward scheme significantly reduces the contribution of noise in auxiliary degrees of freedom to the gravitational wave channel in the frequency range 10--150\,Hz. The typical subtraction factors for Michelson length  noise, signal recycling cavity length noise, and angular noise are 30, 7 and 20, respectively.

\subsection{Oscillator noise}

The RF oscillator used to generate the Pound-Drever-Hall control sidebands has phase and amplitude noise, and these couple to the gravitational wave channel via both sensing intensity noise and displacement noise in the dual-recycled Michelson degrees of freedom.

Noise in the oscillator amplitude causes the RF modulation index to vary with time, thus changing the amount of power contained in the RF sidebands.
Since the total power in the carrier and the RF sidebands is actively controlled, fluctuations in the RF sideband field amplitudes produce fluctuations in the carrier field amplitude (i.e., audio sidebands).
These audio sidebands propagate through the interferometer and couple into the gravitational wave channel via the same mechanisms as laser intensity noise as discussed in Sec.~\ref{noises_laser}.
Additionally, as intensity noise of RF sidebands is not filtered by the common coupled cavity pole and the output mode cleaner has a finite attenuation at the RF sideband frequencies ($\simeq 6\times10^{-5}$\,W/W for the 45\,MHz sidebands), a small amount of sideband power fluctuations appears directly on the GW readout photodiodes.
The oscillator amplitude noise coupling for the 9\,MHz and 45\,MHz sidebands was measured to be 
\begin{equation}
\begin{split}
	L(f) = & 5 \times 10^{-22} \left( \frac{N_{\text{amp}}^9}{10^{-6}} \right) 
	 \frac{1}{K_-(f)}  \frac{\text{m}}{\sqrt{\text{Hz}}} \\
	L(f) = & 5 \times 10^{-21} \left( \frac{N_{\text{amp}}^{45}}{10^{-6}} \right) 
	\frac{1}{K_-(f)} \frac{\text{m}}{\sqrt{\text{Hz}}},
\end{split}
\end{equation}
where $N_{\text{amp}}^9$ and $N_{\text{amp}}^{45}$ is the relative amplitude noise of 9\,MHz and 45\,MHz sidebands in units of $1/\sqrt{\text{Hz}}$.

Oscillator phase noise is converted to RF sideband amplitude noise through any optical path length imbalance in the interferometer's Michelson degree of freedom.
The main sources of imbalance are the intentional asymmetry in the Michelson interferometer and a transmissivity difference of the input test masses (which produces a differential phase delay when the sidebands are reflected from each arm)~\cite{Izumi2015b}. The oscillator phase noise coupling for the 9\,MHz and 45\,MHz sidebands was measured to be
\begin{equation}
\begin{split}
	L(f) = & 10^{-21} \left( \frac{N_{\text{ph}}^9 f}{10^{-2}} \right) 
	\frac{1}{K_-(f)} \frac{\text{m}}{\sqrt{\text{Hz}}} \\
	L(f) = & 10^{-22} \left( \frac{N_{\text{ph}}^{45} f}{10^{-2}} \right) 
	\frac{1}{K_-(f)} \frac{\text{m}}{\sqrt{\text{Hz}}},
\end{split}
\end{equation}
where $N_{\text{ph}}^9$ and $N_{\text{ph}}^{45}$ is the relative phase noise of 9\,MHz and 45\,MHz sidebands in units of $1/\sqrt{\text{Hz}}$.

\subsection{Beam jitter}

\begin{figure}[ht]
	\centering
	\includegraphics[width=8cm]{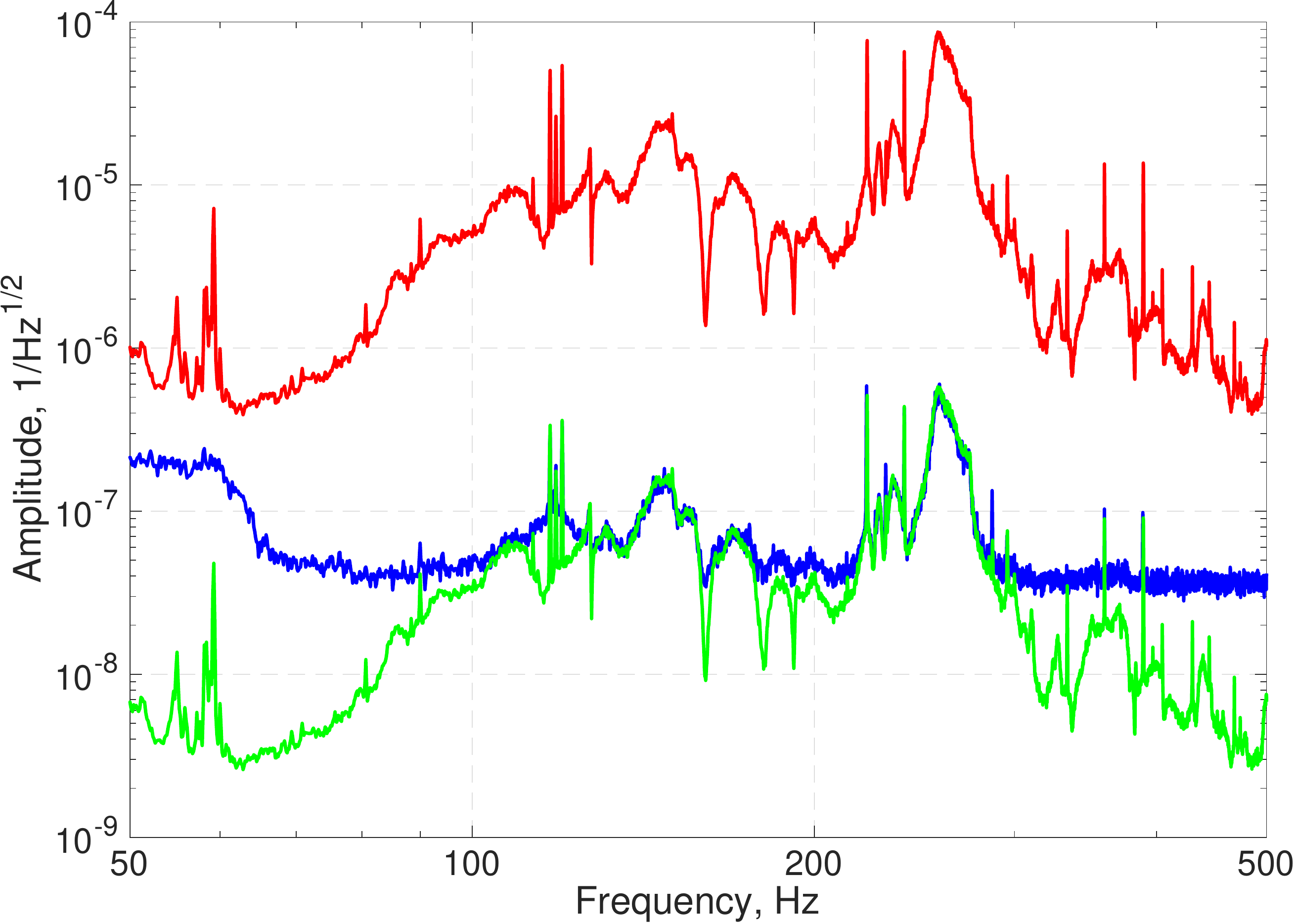}
	\caption{Relative pointing noise before and after the input mode cleaner in L1 interferometer. Acoustic peaks in the L1 and H1 interferometers are at slightly different frequencies.
        The red trace shows the spectrum measured before the input mode cleaner, where laser beam enters the vacuum system.
        The blue trace shows the measured jitter after the input mode cleaner. This measurement is limited by the sensing noise of the quadrant photodetector at a level of $4 \times 10^{-8} ~ /\sqrt{\text{Hz}}$.
	The green trace is the estimated relative pointing noise used in the calculation of the jitter coupling to the gravitational wave channel. This curve is computed by dividing the red spectrum by the filtering coefficient of the input mode cleaner.}
	\label{fig:input_jitter}
\end{figure}

Pointing fluctuations, quantified by the factor $\Delta w/ w$, where $w$ is the beam size and $\Delta w$ is the transverse motion of the beam, are also a source of noise. 
On the input side, significant beam jitter is caused by angular and longitudinal motion of the steering mirrors, located in air. The input mode cleaner, located in vacuum, attenuates the input beam jitter by a factor of $\simeq 150$. Fig. \ref{fig:input_jitter} shows the relative pointing fluctuations before and after the input mode cleaner.

Residual input beam jitter is converted into intensity fluctuations by the interferometer resonant cavities: the power and signal recycling cavities, the arm cavities, and finally the output mode cleaner cavity. In the frequency range 100\,Hz--1\,kHz, the coupling coefficient from relative pointing noise at the interferometer input to the relative intensity noise at the antisymmetric port is $\sim$ 0.01.

Fig.~\ref{fig:data_nb} shows that the contribution of the beam jitter is close to the measured strain noise at a few peaks between 200 and 600\,Hz. These structures in the noise are due to resonances of mirror mounts in the in-air input beam path.
This contribution has been reduced by improving the stiffness of the optical elements, thus reducing the motion.

On the output side, beam jitter is caused by angular motion of the output steering mirrors. These are single pendulum stage suspended optics, located in vacuum. While interferometer alignment is actively controlled to reduce beam jitter, any residual angular motion modulates the power transmitted by the output mode cleaner and thereby couples to the gravitational wave channel.

\subsection{Scattered light noise}

Motion of the suspended optics is significantly reduced compared to the ground, as discussed in Sec.~\ref{seismic_thermal}. However, the vacuum chambers and arm tubes are not isolated from the ground seismic or the ambient acoustic noises. This motion can couple to the gravitational wave channel through scattered light.

\begin{figure}[h]
    \centering
    \includegraphics[width=8.5cm]{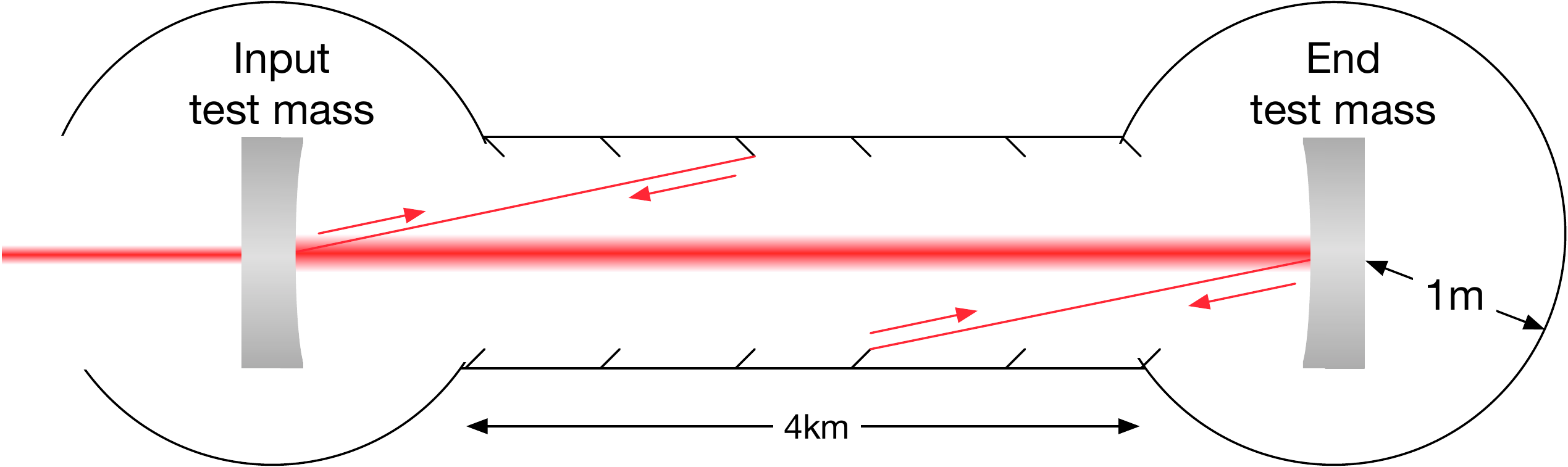}
    \caption{Scattering inside the arm cavity. The test mass coating irregularities and dust determines how much light can be scattered in and out from the main beam. After the scattered light hits the beam tube baffles, which are not isolated from ground motion or acoustic noise, it partially scatters back into the main beam. This process couples motion of beam tubes to the gravitational wave channel.}
    \label{fig:scatter_arm}
    \end{figure}

A small portion of the laser light scatters out of the main beam when it hits the optical components. Part of this light is scattered back from the moving chamber walls, baffles, mirrors, or photodiodes, and couples into the main beam as shown in Fig. \ref{fig:scatter_arm}. Backscattered light modulates the main beam in phase and amplitude, and introduces noise into the gravitational wave channel. The phase modulation is directly detected at the antisymmetric port, and amplitude modulation moves the test masses by means of radiation pressure.
Signifiant scattering processes occur inside the arm cavities, at the input and output ports, and in the recycling cavities.

\subsubsection{Beam tubes}

Light scattered out from the main beam by the test masses couples motion of the 4--km beam tube to the gravitational wave channel. The bi-directional Reflectivity Distribution (BRDF) function of the test masses depends on the imperfections in the mirror surface. If the wavelength of a coating ripple is $\lambda_{r}$, then the angle between the scattered light and the main beam is $\theta = \lambda / \lambda_r$. The amount of power scattered out from the main beam depends on the amplitude of the ripple. The fractional power scattered out in the cone with half angle $\theta \ll 1$ and width $d\theta$ is given by:

\begin{equation}
	\frac{dP_s}{P_{\text{arm}}} \approx \left( \frac{4\pi}{\lambda} \right)^2 
	S \frac{d\theta}{\lambda} = \text{BRDF}_m \times d\Omega
\end{equation}
where $S(\theta/\lambda)=S(\lambda_r^{-1})$ is the power spectral density of the coating aberrations \cite{scatter_peter_hiro, stover_scattering}, and $d\Omega=2 \pi \theta d\theta$ is solid angle of scattering. For $\theta \sim 1$, the BRDF can be approximated as $\text{BRDF}_m = 3 \times 10^{-6} \cos(\theta)$\,$\text{sr}^{-1}$.

Light scattered out from the main beam hits a baffle in the arm tube and scatters back into the main beam. The measured BRDF of the baffle at large angles is $\text{BRDF}_b = 0.02$\,$\text{sr}^{-1}$. In order to get back into the main beam, light from the baffle scatters into the solid angle $\lambda^2/r^2 \times \text{BRDF}_m$ \cite{noise_scatter}, where $r$ is the distance from the baffle to the test mass. The total optical power $P_r$ that recombines with the main beam is determined by the following equation \cite{loss_arm_scatter}:
\begin{equation}
	\frac{dP_r}{P_{\text{arm}}} = \frac{\lambda^2}{r^2}\text{BRDF}_m^2 \text{BRDF}_b d\Omega.
\end{equation}

The coating profiles were measured \cite{design_aligo} and can be approximated as a smooth polynomial function in the wide range of $\lambda_r$ for narrow angle scattering. However, the high-reflectivity coatings applied on the end test masses show a distinct azimuthal
ripple in the coating surface height. The spatial wavelength of the ripple is 7.85\,mm and its maximum amplitude is 1\,nm pk-pk. This ripple is located at radii beyond about 3\,cm from the mirror center and significantly contributes to the scattered light noise \cite{ripple_peter}. The total scattered light noise contribution to the differential arm channel from the tube motion $N_{tube}$ is
\begin{equation}\label{eq:scatter_arm}
	L(f) = \sqrt{2 \frac{\int dP_r}{P_{\text{arm}}}} N_{\text{tube}}(f) \approx 10^{-11} N_{\text{tube}}(f),
\end{equation}
where the integral is computed over all scattering angles (the factor of 2 accounts for the incoherent sum of all four test masses and for the fact that 1/2 of the baffle motion, in power, is in the phase quadrature of the main field). Equation \ref{eq:scatter_arm} accounts only for the phase quadrature and ignores radiation pressure noise. This is a valid assumption for the current optical power $P_{arm} \approx 100$\,kW.

The estimated scattered light noise, coming from the arm cavities, is a factor of 30-100 below the current sensitivity of the interferometer. This result was confirmed by applying periodic mechanical excitation to the beam tube at different frequencies and measuring the response in the gravitational wave channel.

\subsubsection{Vacuum chambers}

\begin{figure}[h]
    \centering
    \includegraphics[width=8.5cm]{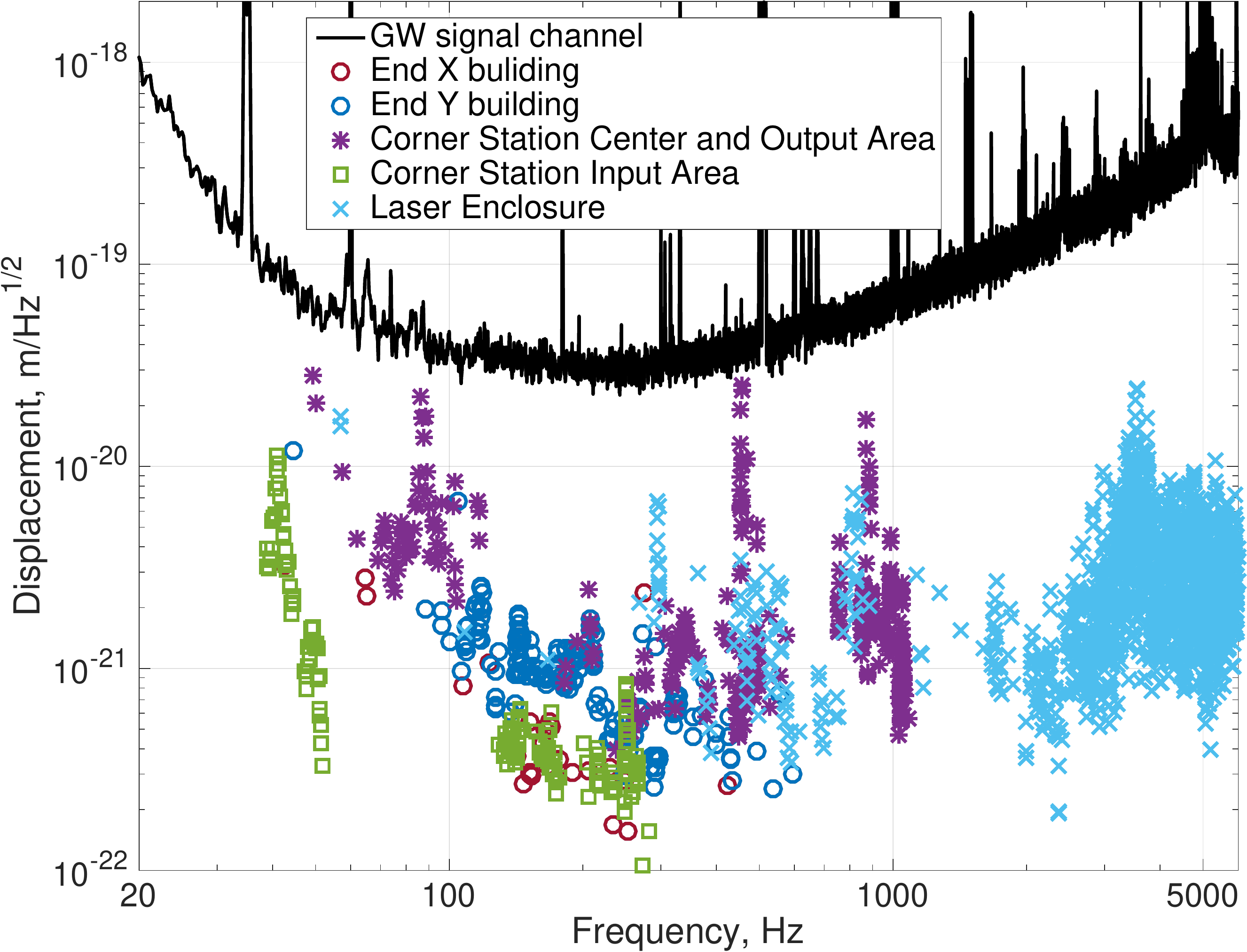}
    \caption{Projected contribution of the ambient acoustic noise to the gravitational wave channel. An acoustic excitation was applied at different locations near the vacuum chambers: near the end test masses of both arm cavities (X and Y), near the dual-recycled Michelson interferometer (corner station), and near the main laser.}
    \label{fig:data_scatter}
\end{figure}

Similar scattering processes occur in the chambers and short tubes in the corner station, where the dual-recycled Michelson interferometer is located. One method to assess the contribution of scattering noise to the detector background is to inject known acoustic signals and measure the response in the gravitational wave channel~\cite{effler_pem}. In general, coupling of scattered light noise is not linear but rather modulated by the low frequency motion of the scattering surfaces. For this reason, instead of measuring the transfer function from the excitation to the sensor, we monitor excess power in the signal spectrum. Then we make a projection of scattered light noise to the gravitational wave channel according to the following equation
\begin{equation}
	L(f) = N_{\text{amb}}(f) \frac{L_{\text{exc}}(f)}{N_{\text{exc}}(f)}, 
\end{equation}
where $L_{\text{exc}}$ and $N_{\text{exc}}$ are the spectra of the gravitational wave channel and of the back scattering element motion, respectively, when an excitation to the element is applied, and $N_{\text{amb}}$ is the motion of the scattering element without any excitation. Fig. \ref{fig:data_scatter} shows that the projected ambient acoustic noise coupling to the gravitational wave channel is below the measured sensitivity. 

\subsubsection{Fringe wrapping}

\begin{figure}[h]
    \centering
    \includegraphics[width=8.5cm]{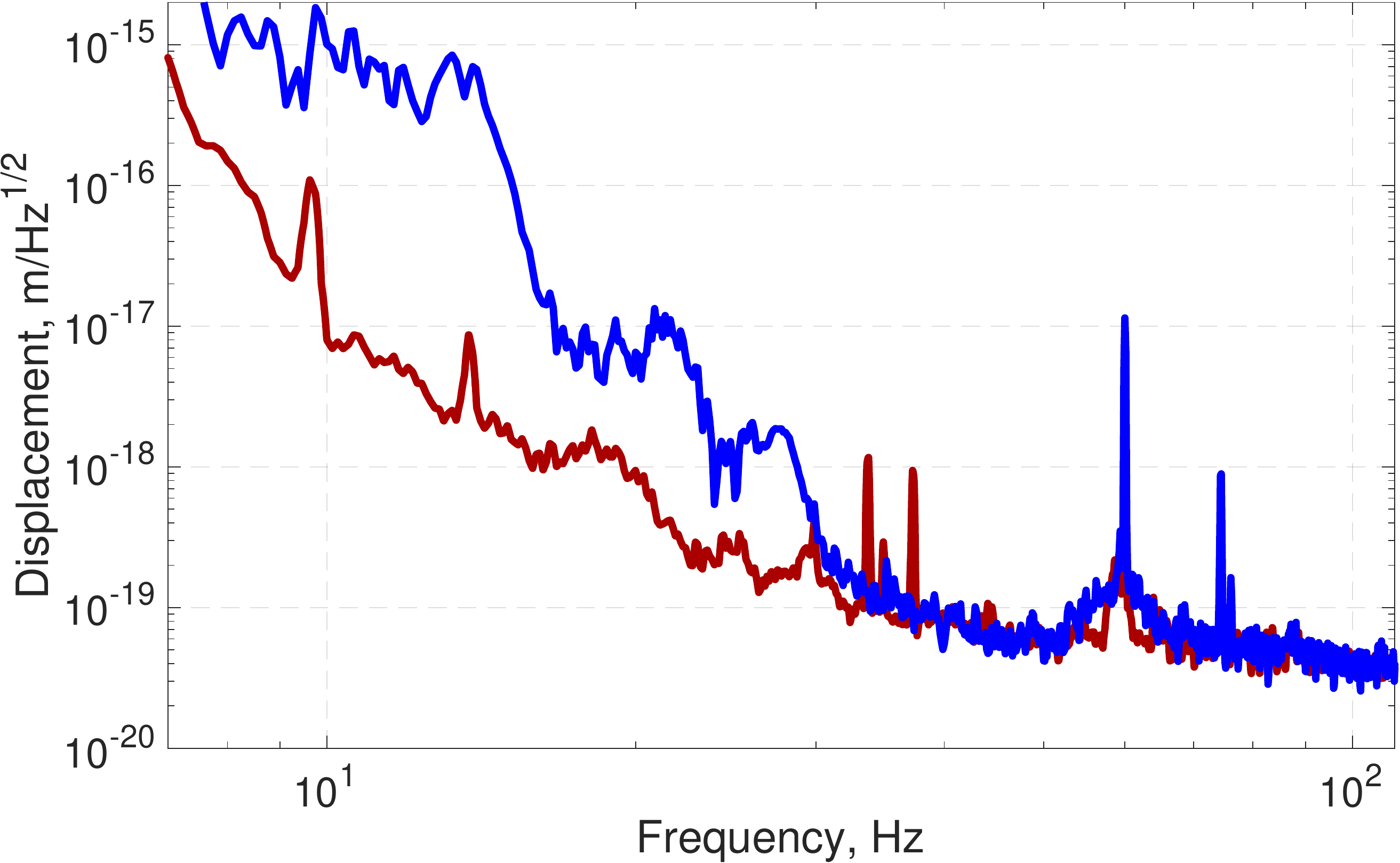}
    \caption{Scattering shelves in the differential arm channel. The red trace shows the spectrum when RMS of the ground velocity is below $\simeq 2$\,um/sec (usual conditions). The blue trace shows the spectrum when the distance between the output mode cleaner and the interferometer was modulated at low frequencies by $\simeq 6$\,um/sec.}
    \label{fig:data_scatter_shelf}
\end{figure}

Scattered light may also manifest itself through up-conversion of the scattering element motion. One example of such a non-linear scattering process is fringe wrapping. In Advanced LIGO fringe wrapping occurs at the antisymmetric port of the interferometer. Optical imperfections in the output mode cleaner cause a fraction of the light ($\sim 1$\,ppm) to travel back into the interferometer. Most of this light is rejected by the output Faraday isolator, but a small fraction of scattered light gets through. Then this light is reflected from the instrument and travels back to the output mode cleaner, with an additional varying phase shift due to the relative motion of the output mode cleaner and the interferometer. The relative intensity fluctuation (RIN) at the output mode cleaner transmission due to backscattering is given by \cite{fringe-wrapping}
\begin{equation}
	\text{RIN}(t) = 2 r \cos (4 \pi N_{\text{omc}}(t)/\lambda),
\end{equation}
where $r=10^{-5}-10^{-4}$ is effective field reflectivity of the interferometer output port and $N_{\text{omc}}(t)$ is the distance fluctuation between the interferometer and the output mode cleaner. Since this distance is not controlled, the amplitude of $N_{\text{omc}}(t)$ can be as large as several wavelengths, and the cosine in the above equation wraps this rapidly varying phase between 0 and $2\pi$, leading to up-conversion of the low frequency motion of the length. The resulting ``scattering shelves'' are seen in the differential arm length spectrum with a cutoff frequency of $2 /\lambda \times dN_{\text{omc}}/dt$. In Advanced LIGO, when the micro-seismic motion is higher than normal, this process increases the gravitational wave channel noise below 20\,Hz~\cite{den_thesis}. Fig.~\ref{fig:data_scatter_shelf} shows scattering shelves in the gravitational wave channel during the low frequency modulation of the distance $N_{\text{omc}}$.

\subsection{Sensing and actuation electronics noise}

This section summarizes noise contributions from electronic circuits in photodetectors, actuators, analog-to-digital (ADC) and digital-to-analog (DAC) converters and whitening boards, all of which are essential for sensing optical signals and actuating on suspensions. From a design perspective, all electronics noise should be smaller than fundamental noises.

For the differential arm length signal, a pair of reverse-biased InGaAs photodiodes, equipped with in-vacuum preamplifiers, measures the light transmitted by the output mode cleaner. 
Subsequently, these signals are acquired by a digital system through analog-to-digital converters, further dividing sensing noise into two types: dark noise and ADC noise.
Dark noise includes any dark current produced by the photodiodes, Johnson-Nyquist noise of the readout transimpedance, and noise in all other downstream analog electronics.
A current noise level of $\sim$10\,$\text{pA}/\sqrt{\text{Hz}}$, at 100\,Hz, is present in each photodetection circuit, equivalent to the shot noise of a DC current of 0.3\,mA. 
This can be compared against the actual operating current of 10\,mA.
Taking the coherent sum of two photodetectors into account, we estimate the dark noise to be a factor of 8.2 lower than the shot noise at 100\,Hz, as shown in Fig. \ref{fig:data_nb}. ADC input noise is suppressed by inserting additional analog gain and filtering, referred to as ``whitening filters''. 
An offline measurement of the ADC noise shows that it is below the current best noise level by a factor of more than 10 over the entire measurement frequency band.

The other important noise in this category is noise in the actuation used to apply feedback control forces on the mirrors. 
Any excess noise at the level of the required actuation couples directly to mirror displacement. The most critical actuation noise is due to the digital-to-analog convertors that bridge the digital real-time control process and the analog suspension drive electronics. 
It is a significant challenge to achieve both the high-range actuation, needed to bring the interferometer into the linear regime from an uncontrolled state (lock acquisition) \cite{lock_aligo_als}, and low-noise actuation for operation in the observation state. 
This issue has been tackled by installing a gain-switchable force controller, which has several operational states. After the interferometer is brought into the linear regime, the controller state is changed from the high-dynamic-range to the low-noise state. 
Also, noise from the digital-to-analog converter is mechanically filtered via the suspension force-to-displacement transfer function above $\sim 0.5$\,Hz. The current estimate puts the actuation noise is as low as $3\times 10^{-18}~\text{m} /\sqrt{\text{Hz}}$ at 10\,Hz.

Lastly, active damping of the suspension systems is known to introduce noise. Below 5\,Hz, the high-Q suspension resonances are damped by sensing the motion of the suspension relative to its support using shadow sensors \cite{bosem_garbone}. 
According to dynamical suspension models, noise from the local damping control is estimated to be $2\times10^{-18}~\text{m}/\sqrt{\text{Hz}}$ at 10\,Hz, and rapidly decreases at higher frequencies.

\subsection{Noise stationarity}

\begin{figure}[ht]
    \centering
    \includegraphics[width=8cm]{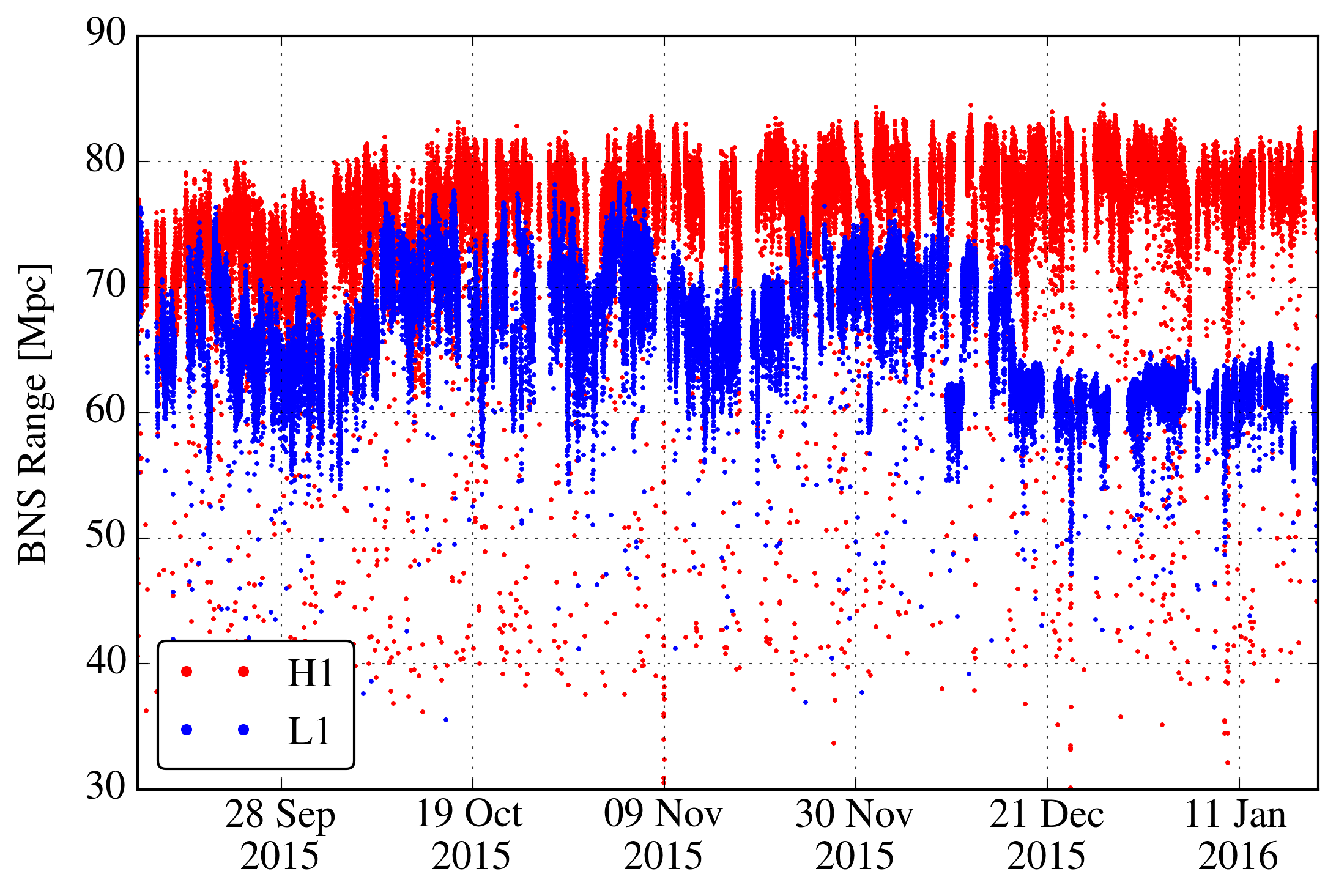}
    \caption {Sensitivity of the two Advanced LIGO detectors to binary neutron star inspirals, averaged over sky position and orientation and 1 minute of data. The sensitivity drop in the L1 interferometer at the end of the run was caused by electronics noise at one of the end stations. This noise was identified and eliminated shortly after the observing run.}
    \label{fig:data_range}
\end{figure}

A common figure of merit for ground-based interferometric detectors is their sensitivity to the inspiral of two neutron stars, averaged over relative orientations of the binary system and sky locations.
A plot of the sensitivity of the LIGO detectors to signals of this type is given in Fig.~\ref{fig:data_range}, over a one-month timescale.

\begin{figure}[ht]
        \centering
        \includegraphics[width=8cm]{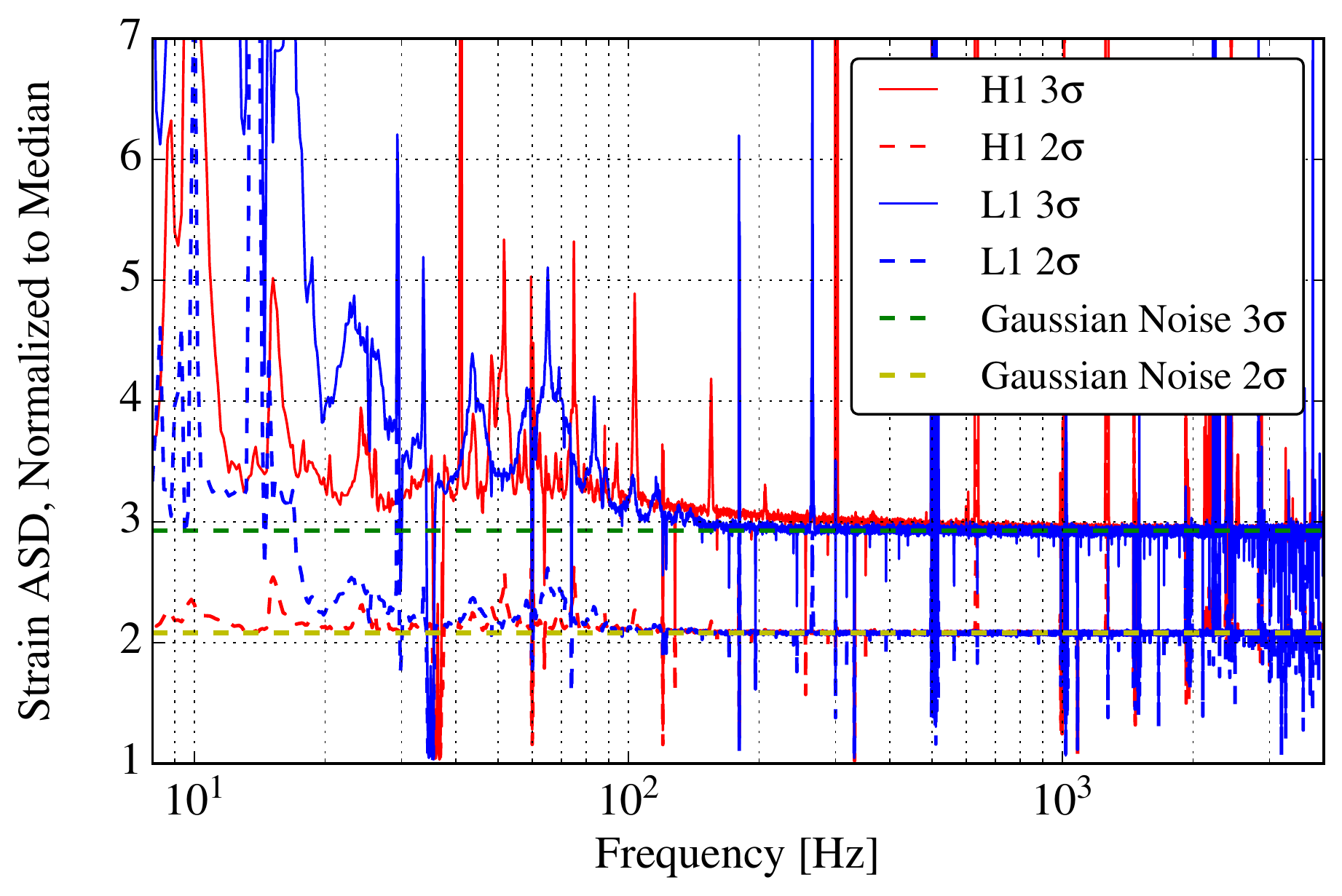}
        \caption {95th ($2\sigma$) and 99.7th ($3\sigma$) percentiles of the noise in each frequency bin during five days of operations (Oct 15-20). The expectations for stationary Gaussian noise at each percentile are given by the colored dashed lines. Both detectors exhibit some non-stationary behavior at low frequencies, due to varying noise couplings with the environment.}
        \label{fig:data_rayleigh}
\end{figure}

If the sum of all the noises is truly Gaussian and stationary, the strain noise density at a given frequency will vary randomly in time following a Rayleigh distribution.
In Fig.~\ref{fig:data_rayleigh}, we compare the 95th and 99.7th percentiles of the noise in each frequency bin to the expectation for stationary Gaussian noise.
Deviations from the expectation are due to non-stationary noises (e.g., transient environmental disturbances that generate very short-duration bursts of excess noise, mostly at low frequency) or narrowband features that are coherent over long timescales.
Above 100\,Hz, the deviation from stationary Gaussian noise is small. Below 100\,Hz, the fluctuations can mask or mimic gravitational waves and must be addressed by further commissioning and, for O1 data, through vetoes that are applied following data collection and analysis.

The characterization and mitigation of the detector noise is the focus of a large collaborative effort between instrument specialists and the gravitational wave data analysis community \cite{data_detchar, jess_detchar_2016}.

\section{Conclusions}
\label{sec:conclusions}

The first Advanced LIGO observational run (O1) started in September 2015 and concluded in January 2016. The observatory was running at unprecedented sensitivity to gravitational waves in the frequency range 10\,Hz--10\,kHz. The average distance at which Advanced LIGO could detect the coalescence of binary black hole systems with individual masses of 30\,$M_\odot$ and with signal-to-noise ratio of 8 was 1.3\,Gpc. The reach for binary neutron star inspirals during the first science run was about 75\,Mpc.

The commissioning of Advanced LIGO lasted for $\sim$ 1 year before the beginning of O1. During this period, a variety of technical noise sources was discovered and eliminated. In this paper, we discussed the dominant noise sources that limited Advanced LIGO sensitivity during the first science run. The coupling of auxiliary degrees of freedom, laser amplitude noise, suspension actuation and other technical noises considered in this paper were significantly reduced. 

Future work is required to find the remaining noise sources. In particular, below 100\,Hz, the sum of all known noise sources in the gravitational wave channel could not explain the measured sensitivity curve. 

Above 100\,Hz, the Advanced LIGO sensitivity was limited mostly by photon shot noise. For this reason, one certain activity on the commissioning agenda is to increase the interferometer input power and ultimately to introduce squeezed states of light \cite{lisa_squeezed, hartmut_squeezed, geo_squeezed}. During the first science run, Advanced LIGO operated in the low power regime: input power was 25\,W out of maximum laser power of $\simeq 180$\,W. A set of technical difficulties must be overcome before power can be increased. First of all, parametric instabilities~\cite{data_pi_llo}, which arise at high power, should be damped to keep the interferometer in the linear regime. Second, angular instabilities in the arm cavities~\cite{ang_instability} are expected to occur when the circulating arm power reaches $\simeq 500$\,kW. This problem will be addressed by changing the angular control system control topology. Lastly, power levels on the photodetectors should be adjusted in order to avoid their damage during lock losses, when stored optical energy leaves the interferometer through the output ports.

\section*{Acknowledgments}

The authors gratefully acknowledge the support of the United States
National Science Foundation (NSF) for the construction and operation of the
LIGO Laboratory and Advanced LIGO as well as the Science and Technology Facilities Council (STFC) of the
United Kingdom, the Max-Planck-Society (MPS), and the State of
Niedersachsen/Germany for support of the construction of Advanced LIGO
and construction and operation of the GEO600 detector.
Additional support for Advanced LIGO was provided by the Australian Research Council.
The authors gratefully acknowledge the Italian Istituto Nazionale di Fisica Nucleare (INFN),
the French Centre National de la Recherche Scientifique (CNRS) and
the Foundation for Fundamental Research on Matter supported by the Netherlands Organisation for Scientific Research,
for the construction and operation of the Virgo detector
and the creation and support  of the EGO consortium.
The authors also gratefully acknowledge research support from these agencies as well as by
the Council of Scientific and Industrial Research of India,
Department of Science and Technology, India,
Science \& Engineering Research Board (SERB), India,
Ministry of Human Resource Development, India,
the Spanish Ministerio de Econom\'ia y Competitividad,
the Conselleria d'Economia i Competitivitat and Conselleria d'Educaci\'o, Cultura i Universitats of the Govern de les Illes Balears,
the National Science Centre of Poland,
the European Commission,
the Royal Society,
the Scottish Funding Council,
the Scottish Universities Physics Alliance,
the Hungarian Scientific Research Fund (OTKA),
the Lyon Institute of Origins (LIO),
the National Research Foundation of Korea,
Industry Canada and the Province of Ontario through the Ministry of Economic Development and Innovation,
the Natural Science and Engineering Research Council Canada,
Canadian Institute for Advanced Research,
the Brazilian Ministry of Science, Technology, and Innovation,
Russian Foundation for Basic Research,
the Leverhulme Trust,
the Research Corporation,
Ministry of Science and Technology (MOST), Taiwan
and
the Kavli Foundation.
The authors gratefully acknowledge the support of the NSF, STFC, MPS, INFN, CNRS and the
State of Niedersachsen/Germany for provision of computational resources.

\bibliography{paper}

\end{document}